\documentclass[11pt, nofootinbib, notitlepage]{revtex4-1}

\usepackage{braket}
\usepackage{enumerate} 
\usepackage{amsmath,amsfonts, amssymb, amsthm, dsfont}
\usepackage{graphicx,setspace}

\usepackage{yfonts,bbm}
\usepackage{bm,breqn}
\usepackage{mathrsfs}
\usepackage{graphicx}
\usepackage[caption=false]{subfig}
\usepackage{verbatim}

\usepackage[pdftex,colorlinks=true,linkcolor=darkblue,citecolor=blue,urlcolor=darkred]{hyperref}
\usepackage{xcolor}
\usepackage{xspace}
\definecolor{darkblue}{rgb}{0.,0.,0.4}
\definecolor{darkred}{rgb}{0.5,0.,0.}

\newcommand{\bi}{\begin{itemize}}
	\newcommand{\ei}{\end{itemize}}
\newcommand{\beq}{\begin{equation}}
\newcommand{\eeq}{\end{equation}}
\newcommand{\bea}{\begin{eqnarray}}
\newcommand{\eea}{\end{eqnarray}}
\newcommand{\bpm}{\begin{pmatrix}}
\newcommand{\epm}{\end{pmatrix}}
\newtheorem{theorem}{Theorem}
\newtheorem{claim}{Claim}
\newtheorem{lemma}{Lemma}

\newlength{\fighskip} \fighskip=2pt
\newlength{\figvskip} \figvskip=3pt

\newcommand*{\figbox}[2]{{
  \def\figscale{#1}
  \def\arraystretch{0.8}
  \arraycolsep=0pt
  \begin{array}{c}
    \vbox{\vskip\figscale\figvskip
      \hbox{\hskip\figscale\fighskip
        \includegraphics[scale=\figscale]{#2}}}
  \end{array}}}

\begin{document}

\title{Randomly Monitored Quantum Codes}
\author{Dongjin Lee}
\author{Beni Yoshida} 
\affiliation{\PI}

\newcommand*{\PI}{Perimeter Institute for Theoretical Physics, Waterloo, Ontario N2L 2Y5, Canada}
\newcommand*{\UW}{Department of Physics and Astronomy, University of Waterloo, Waterloo, Ontario N2L 3G1, Canada}

\begin{abstract}
Quantum measurement has conventionally been regarded as the final step in quantum information processing, which is essential for reading out the processed information but collapses the quantum state into a classical state. 
However, recent studies have shown that quantum measurement itself can induce novel quantum phenomena. One seminal example is a monitored random circuit, which can generate long-range entanglement faster than a random unitary circuit. Inspired by these results, in this paper, we address the following question: When quantum information is encoded in a quantum error-correcting code, how many physical qubits should be randomly measured to destroy the encoded information? We investigate this question for
various quantum error-correcting codes and derive the necessary and sufficient conditions for destroying the information through measurements. 
In particular, we demonstrate that for a large class of quantum error-correcitng codes, it is impossible to destroy the encoded information through random single-qubit Pauli measurements when a tiny portion of physical qubits is still unmeasured. 
Our results not only reveal the extraordinary robustness of quantum codes under measurement decoherence, but also suggest potential applications in quantum information processing tasks. 
\end{abstract}     

\maketitle


\section{Introduction}

One of the most well-known mysteries in modern physics is that we perceive the world as a classical object despite that the underlying laws of physics are supposedly quantum.
A popular anecdote, demonstrating this tension, is the Schr\"{o}dinger's cat which considers a macroscopic superposition of a dead and an alive cat. 
While it is mind-boggling that such a state is in principle allowed to exist, this puzzle can be potentially resolved by noting that the information about whether the cat is dead or alive can be learned easily by simply measuring the cat, namely  
\begin{align}
\alpha|\mathrm{dead}\rangle + \beta |\mathrm{alive}\rangle
\xrightarrow{\;\; \text{measurement} \;\; } |\mathrm{dead}\rangle \ \text{or} \ |\mathrm{alive}\rangle,
\end{align}
implying its instability.
There is an information theoretic interpretation of the Schr\"{o}dinger's cat.
Imagine that we have a qubit, prepared in $\alpha|0\rangle_{\text{dead}} + \beta|1\rangle_{\text{alive}}$, and encode it into 
\begin{align}
\alpha|0\rangle_{\text{dead}} + \beta|1\rangle_{\text{alive}}\xrightarrow{\;\; \text{encode} \;\; } 
\alpha|\mathrm{dead}\rangle + \beta |\mathrm{alive}\rangle.
\end{align}
That the Schr\"{o}dinger's cat easily decoheres can be understood as the loss of encoded quantum information due to measurement.
Indeed, an analogue of this encoding, often called the cat code, considers an encoding via
$|0\rangle_{\text{dead}} \longrightarrow |0\rangle^{\otimes n}, \ |1\rangle_{\text{alive}} \longrightarrow |1\rangle^{\otimes n}$. 
When a single $Z$ measurement is performed, the state collapses into either dead or alive state
\bea
\alpha |0\rangle^{\otimes n} + \beta |1\rangle^{\otimes n}
\xrightarrow{\;\; Z_1 \text{measurement} \;\; } 
|0\rangle^{\otimes n} \text{ or } |1\rangle^{\otimes n} 
\eea
and the original encoded quantum information is lost.
These observations are often acknowledged as the reason why macroscopic entanglement does not exist in our world.
A conventional wisdom is that measurements destroy quantum entanglement due to collapse of wavefunctions. 

Recent developments on studies of monitored hybrid quantum circuits, however, provide an interesting new perspective on this problem, suggesting that quantum entanglement and encoded information may tolerate significant amount of decoherence from measurements (see~\cite{Skinner:2019aa, Li:2018aa, Chan:2019, Choi:2020aa, Li2021, Li:2021aa, Bao2020, PhysRevLett.125.070606, PhysRevB.100.134306, PhysRevX.10.041020} for an incomplete list of references, and see~\cite{review} for a review). 
In particular, it has been found that the encoded quantum information, prepared by some short-depth quantum circuits, may persist even when an overwhelming portion of qubits are projectively measured~\cite{Choi:2020aa}. 
Evidences suggesting a similar phenomena are also found in the AdS/CFT correspondence~\cite{yoshida2022projective, Antonini_2022}.
These findings motivate us to revisit the century-old problem concerning the fate of quantum information under projective measurements.
In this paper, we ask whether the encoded information in a quantum error-correcting code (QECC) is retained when a part of the system is measured in locally random basis. 
Our main finding is that QECCs possess extraordinary robustness against decoherence due to projective measurements. 
Specifically, we will demonstrate that many examples of QECCs achieve the maximal measurement threshold $p_m^{th}=1$ under uniformly random local measurements. 
This means that, even when 99.99\% of the qubits are projectively measured, the remaining 0.01\% of qubits will still retain the encoded information!
Examples of such codes include concatenated five-qubit and seven-qubit codes, 2D topological codes, as well as holographic QECCs. 

To be precise, suppose that $k$ logical qubits are encoded into $n$ physical qubits in a QECC. 
We measure each physical qubit in local random Pauli basis (i.e. in the $X$,$Y$, or $Z$ basis) with probability $(p_X,p_Y,p_Z)$. 
Let $p_m \equiv p_X + p_Y + p_Z <1$, meaning that $(1-p_m)n$ physical qubits will remain untouched. 
We fix the relative measurement frequencies as $(\alpha_X,\alpha_Y,\alpha_Z)$ with $\alpha_X + \alpha_Y + \alpha_Z =1$ so that $(p_X,p_Y,p_Z)=(p_m \alpha_X, p_m \alpha_Y, p_m \alpha_Z)$. We think of increasing $p_m$ and hope to find the measurement threshold $p_m^{th}$ below which the encoded information remains recoverable with high probability. 
The central result of this paper is an observation that a large class of QECCs achieve $p_m^{th}=1$ for some relative measurement frequencies $(\alpha_X,\alpha_Y,\alpha_Z)$.
We studied four different families of QECCs, 1) concatenated QECCs, 2) 2D topological QECCs, 3) Haar random code, and 4) holographic QECCs. 
In all the examples, we confirmed $p_m^{th}=1$ for some relative measurement frequencies $(\alpha_X,\alpha_Y,\alpha_Z)$. 

This paper is organized as follows. 
In section~\ref{sec:setup}, we begin by presenting several generic results and observations concerning measurements in QECCs.
In section~\ref{sec:concatenation}, we study several examples of concatenated quantum codes.
In section~\ref{sec:2D}, we study the 2D toric code, the 2D color code, and the 2D Bacon-Shor code. 
In section~\ref{sec:Haar}, we discuss the effect of local random measurements on Haar random code, and also the effect of Haar random measurements on a QECC, as opposted to local random measurements. 
In section~\ref{sec:QG}, we briefly discuss the holographic quantum codes. 
We conclude the paper in section~\ref{sec:outlook}. 

\emph{Note added:} After completion of this work, we became aware of~\cite{Botzung:2023} which considers a similar problem.

\section{Some theorems about monitored quantum codes}\label{sec:setup}

In this section, we present some generic results regarding projective measurements for stabilizer and subsystem codes. 

\subsection{Stabilizer update rule}

We begin by discussing how the stabilizer group is updated after projective measurements. 
Suppose that the initial state $|\psi\rangle$ is stabilized by the stabilizer group $\mathcal{S}$, namely $U |\psi\rangle = + |\psi\rangle$ for $U \in \mathcal{S}$.
Projective measurements are performed on a set of mutually commuting independent Pauli operators $\{P_{1},\cdots, P_m\}$. 
For discussions in this section, $P_{j}$'s do not need to be single-qubit Pauli operators. 
The goal is to find the \emph{updated stabilizer group} $\mathcal{S}'$ after projective measurements. 
Letting $m_j=\pm 1$ be the measurement outcomes of $P_{j}$, define measured Pauli operators $M_{j}$ by
\begin{align}
M_{j} \equiv m_{j}P_{j}
\end{align}
so that the measurement outcome of $M_{j}$ is $+1$. 
Note that the values of $m_j$ are not arbitrary and must be consistent with the stabilizer condition. 
Up to a normalization factor, the post measurement state can be expressed as
\begin{align}
|\psi'\rangle \propto \prod_{j}(I+M_j) | \psi\rangle.
\end{align}
Define the \emph{measured Pauli operator group} $\mathcal{M}$ by
\begin{align}
\mathcal{M} \equiv \langle M_{1},\cdots, M_m \rangle
\end{align}
and the centralizer group $\mathcal{C}(\mathcal{M})$, consisting of Pauli operators that commute with $M_{j}$'s, by
\begin{align}
\mathcal{C}(\mathcal{M}) \equiv \big\langle U \in \mathrm{Pauli}  :  [U,\mathcal{M}]=0 \big\rangle. 
\end{align}

\begin{theorem}\label{theorem:update}
The updated stabilizer group $\mathcal{S}'$ is given by
\begin{align}
\mathcal{S}' = \big\langle \mathcal{S}\cap \mathcal{C}(\mathcal{M}) , \ \mathcal{M} \big\rangle.
\end{align}
\end{theorem}

A version of this statement can be found in~\cite{Nielsen_Chuang}, and we present a proof for the sake of completeness. Some readers might be worried about the potential sign issue in considering $\mathcal{S}\cap \mathcal{C}(\mathcal{M})$. 
We constructed $\mathcal{M}$ by using physically possible measurement outcomes $m_j$, and its sign structure is consistent with $\mathcal{S}$. 

\begin{proof}
Recall that the initial state $|\psi\rangle\langle \psi|$ can be expressed as $|\psi\rangle\langle \psi| \propto \sum_{U \in \mathcal{S}} U$ 
up to an overall normalization. The state after the measurement is given by 
\begin{align}
|\psi'\rangle\langle \psi'|
\propto \sum_{M, M' \in \mathcal{M}} M |\psi\rangle\langle \psi| M'  
\propto \sum_{M, M' \in \mathcal{M}}\sum_{U\in \mathcal{S}} M U M'.
\end{align}
Using the following relation
\begin{align}
\sum_{M, M' \in \mathcal{M}} M U M' = 0 \qquad \big(U \not\in \mathcal{C}(\mathcal{M})\big),
\end{align}
one obtains 
\begin{equation}
\begin{split}
|\psi'\rangle\langle \psi'|
\propto 
&\sum_{M, M' \in \mathcal{M}}\sum_{U \in \mathcal{S}\cap \mathcal{C}(\mathcal{M})}  MUM'   
\propto \sum_{M \in \mathcal{M}}\sum_{U \in \mathcal{S}\cap \mathcal{C}(\mathcal{M})}  MU  
\propto \sum_{U \in \langle \mathcal{S}\cap \mathcal{C}(\mathcal{M}), \ \mathcal{M} \rangle } U.
\end{split}
\end{equation}
Hence we have $\mathcal{S}' = \langle \mathcal{S}\cap \mathcal{C}(\mathcal{M}), \ \mathcal{M} \rangle$. 
\end{proof}

This proof can be straightforwardly generalized to the case where the initial state $\rho$ is a \emph{stabilizer mixed state} since
the maximally mixed state $\rho$, stabilized by $\mathcal{S}$, can be written as $\rho \propto \sum_{U \in \mathcal{S}} U$.
\subsection{Information preservation for stabilizer codes}

Next, we study the information preservation conditions. 
Suppose that $k$ logical qubits are encoded into $n$ physical qubits in a stabilizer code, and the code is projectively measured by $\{M_{1},\cdots, M_m\}$. 
We present the necessary and sufficient conditions for the information preservation under projective measurements for stabilizer codes. 

\textbf{Choi state:}
Let $\Xi$ be an encoding isometry mapping an input $k$-qubit state $|\psi\rangle$ into an encoded $n$-qubit state $|\widetilde{\psi}\rangle$. 
The \emph{Choi state} is constructed by applying $\Xi$ to $k$ copies of EPR pairs:
\begin{align}
|\Phi\rangle \equiv (\Xi \otimes I) |\mathrm{EPR}\rangle^{\otimes k} = 
\figbox{0.5}{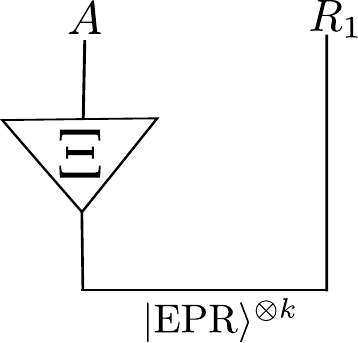}
\end{align}
where the Choi state $|\Phi\rangle$ is an $(n+k)$-qubit state supported on the system $A$ and the reference $R$.
In terms of stabilizer generators, the Choi state can be characterized as follows. Let $\mathcal{S}_{\mathrm{code}}$ be the stabilizer group of the original code. 
The stabilizer group $\mathcal{S}_{\mathrm{Choi}}$ for the Choi state $|\Phi\rangle$ is
\begin{align}
\mathcal{S}_{\mathrm{Choi}} = \Big\langle \mathcal{S}_{\mathrm{code}}\otimes I, \{\bar{X}_{j} \otimes X_{j}, \bar{Z}_{j} \otimes Z_{j}\}_{j=1,\cdots,k}  \Big\rangle \label{eq:Choi_stab}
\end{align} 
where $\bar{X}_{j}, \bar{Z}_{j}$ are logical $X_{j}, Z_{j}$ operators. 

That $k$ logical qubits are encoded in an $n$ physical qubits can be seen from that the system $A$ and the reference $R$ are maximally entangled in the Choi state $|\Phi\rangle$. Namely, the mutual information is maximal;
$I(A,R) \equiv S_A + S_R - S_{AR} = 2k$. 
The condition to retain the encoded information after projective measurement can be expressed concisely by using the updated Choi state.

\begin{lemma} 
The encoded information in a stabilizer code is retained after projective measurements if and only if 
\begin{align}
I(A,R)=2k
\end{align}
in the post measurement Choi state $|\Phi'\rangle$.
\end{lemma}

It is useful to represent the above condition $I(A,R)=2k$ in terms of stabilizer generators. 
Since the post measurement state is pure, the condition is equivalent to $S_R=k$. 
Using the entropy formula for stabilizer states~\cite{Fattal04}, this condition is equivalent to that there is no non-trivial stabilizer operator supported on $R$.

\begin{lemma}
The encoded information in a stabilizer code is retained after projective measurements if and only if, in the Choi state, there is no non-trivial stabilizer operator supported on $R$ in the updated stabilizer group $\mathcal{S}'_{\mathrm{Choi}}$.
\end{lemma}

\textbf{Cleaning lemma for measurement:}
Define a group of Pauli logical operators by
\begin{align}
\mathcal{L} \equiv \big\langle U \in \mathrm{Pauli} : [U,\mathcal{S}]=0 \big\rangle.
\end{align}
Note that stabilizer operators can be viewed as trivial logical operators. 


\begin{theorem}\label{theorem:stabilizer}
The following three statements, concerning Pauli measurements on a stabilizer code, are equivalent: 
\begin{enumerate}[i)]
\item The measured Pauli operator group $\mathcal{M}$ does not contain non-trivial logical operator, namely
\begin{align}
\mathcal{M}\cap \mathcal{L} \subseteq \mathcal{S}.
\end{align}
\item For any non-trivial logical operator $\ell$, there always exists an equivalent logical operator $\ell' \sim \ell$ that commutes with $\mathcal{M}$, namely $\ell'\in \mathcal{C}(\mathcal{M})$.
\item The stabilizer code retains the encoded information after the measurements with respect to $\{M_{1},\cdots, M_m\}$. Namely, in the Choi state, 
\begin{align}
I(R,A) = 2k.
\end{align}
\end{enumerate}
\end{theorem}

Intuitively, the statement i) says that projective measurement by $\mathcal{M}$ should not reveal any encoded information of the code at all, and the statement ii) says that encoded logical qubits must be immune to projective measurements by having logical operators which commute with $\mathcal{M}$. 
These two statements are equivalent, and are the necessary and sufficient conditions for the information preservation. The proof is presented in appendix~\ref{app:proof}.

This theorem can be interpreted as a generalization of the \emph{cleaning lemma}~\cite{Bravyi09}. 
Let us recall that, for stabilizer codes, the cleaning lemma says that the following two statements are equivalent.

\begin{enumerate}[a)]
\item A subset of qubits, denoted by $A$, is correctable, meaning that no logical operator can be supported on $A$. 
In other words, the code can tolerate the erasure error which removes qubits on $A$.

\item For any non-trivial logical operator $\ell$, one can always find an equivalent expression $\ell'$ which is fully supported on the complement $A^c$. 
In other words, one can append suitable stabilizer operator $U$ so that $\ell'=\ell U$ is supported exclusively on $A^c$. 
\end{enumerate}

Returning to the measurement cases, let us specialize in the cases where $\mathcal{M}$ consists of projective measurements on each qubit in a subset $A$. 
Combining the statement i) and ii) from the theorem, we arrive at the following result.

\begin{lemma}
The following two statements are equivalent:
\begin{enumerate}[a)]
\item The stabilizer code retains the encoded information after the projective measurements that acts on qubits on $A$ with respect to $\mathcal{M}$.  

\item For any non-trivial logical operator $\ell$, 
there always exists an equivalent expression logical operator $\ell'\sim \ell$ which can be written as
\begin{align}
\ell' = \ell'_{A^c} \otimes \ell'_{A}, \qquad \text{where} \quad \ell'_{A^c}\not=I_{A^c}, \quad \ell'_{A} \in \mathcal{M}.
\end{align}
Furthermore, $\ell'_{A^c} \otimes I_{A}$ is a logical operator for the post measurement code. 
\end{enumerate}
\end{lemma}

\subsection{Information preservation for subsystem codes}

In this subsection, we extend the above result to subsystem codes~\cite{Poulin_2005}. 

\textbf{Subsystem code:}
Let $\mathcal{S}$ and $\mathcal{G}$ be the stabilizer and gauge groups of the subsystem code respectively. 
Let $S_{1},\cdots, S_{n-k-g}$ be independent stabilizer generators of $\mathcal{S}$.
Consider the subspace stabilized by $\mathcal{S}$;
$ 
\mathcal{H}_{\mathrm{sub}} = \big\{|\psi\rangle : S_{j}|\psi\rangle = + |\psi\rangle \big\} 
$ 
which contains $g+k$ logical qubits.
The central idea of subsystem codes is to use the $k$ \emph{bare} logical qubits only while ignoring $g$ \emph{gauge} logical qubits associated with the gauge group $\mathcal{G}$. 
The total Hilbert space can be written as a direct sum $\mathcal{H} = \bigoplus_{s} \mathcal{H}^{(s)}$ 
where $s$ represents a $(n-k-g)$-component vector which records eigenvalues of $S_{j}$, and the code subspace is given by $\mathcal{H}_{\mathrm{sub}} = \mathcal{H}^{(\vec{1})}$ with $s=\vec{1}$.
The Hilbert space further factorizes as follows~\cite{Bacon06}:
\begin{align}
\mathcal{H}  = \bigoplus_{s} \mathcal{H}^{(s)}_{\mathrm{gauge}}\otimes \mathcal{H}^{(s)}_{\mathrm{bare}} \label{eq:decomposition}
\end{align}
where the action of a gauge operator $U_{\mathrm{gauge}}\in \mathcal{G}$ has the following form: 
\begin{align}
U_{\mathrm{gauge}} = \bigoplus_{s} U^{(s)}_{\mathrm{gauge}} \otimes I^{(s)}_{\mathrm{bare}}.
\end{align}
The stabilized code subspace $\mathcal{H}_{\mathrm{sub}}$ can be factorized as 
$\mathcal{H}_{\mathrm{sub}} = \mathcal{H}^{(\vec{1})} = \mathcal{H}^{(\vec{1})}_{\mathrm{gauge}}\otimes \mathcal{H}^{(\vec{1})}_{\mathrm{bare}}
$
where $\mathcal{H}^{(\vec{1})}_{\mathrm{gauge}}$ and $\mathcal{H}^{(\vec{1})}_{\mathrm{bare}}$ support gauge and bare logical qubits respectively~\footnote{The Hilbert space decomposition in Eq.~\eqref{eq:decomposition} can be constructed by choosing one particular set of bare and gauge logical operators and defining the basis states by their eigenvectors.}.

Following~\cite{Bravyi2011}, a group of \emph{bare} logical operators is defined by
\begin{align}
\mathcal{L}_{\mathrm{bare}} \equiv \big\langle U \in \mathrm{Pauli} : [U,\mathcal{G}]=0 \big\rangle
\end{align}
and its action has the following form
\begin{align}
U_{\mathrm{bare}} = \bigoplus_{s} I^{(s)}_{\mathrm{gauge}} \otimes U^{(s)}_{\mathrm{bare}}.
\end{align}
Non-trivial bare Pauli logical operators must satisfy $U^{(\vec{1})}_{\mathrm{bare}}\not\propto I$. 
Two bare logical operators are said to be equivalent when $\ell_{\mathrm{bare}} \ell_{\mathrm{bare}}' \in \mathcal{S}$.
A group of \emph{dressed} logical operators is defined by 
\begin{align}
\mathcal{L}_{\mathrm{dressed}} \equiv \big\langle U \in \mathrm{Pauli} : [U,\mathcal{S}]=0 \big\rangle
\end{align}
and its action has the following form
\begin{align}
U_{\mathrm{dressed}} = \bigoplus_{s} V^{(s)}_{\mathrm{gauge}} \otimes U^{(s)}_{\mathrm{bare}}. 
\end{align}
Non-trivial dressed Pauli logical operators must satisfy $U^{(\vec{1})}_{\mathrm{bare}}\not\propto I$.   
Two dressed logical operators are said to be equivalent when $\ell_{\mathrm{dressed}} \ell_{\mathrm{dressed}}' \in \mathcal{G}$.

\textbf{Choi state:}
Let $\Xi$ be an encoding isometry that maps an input $g+k$-qubit state into an encoded $n$-qubit state. 
The Choi state of a subsystem code can be constructed by applying the isometry $\Xi$ onto $g+k$ copies of EPR pairs:
\begin{align}
|\Phi\rangle = (\Xi \otimes I_{\mathrm{gauge}} \otimes I_{\mathrm{bare}}) |\mathrm{EPR}\rangle^{\otimes g+k} = \figbox{0.5}{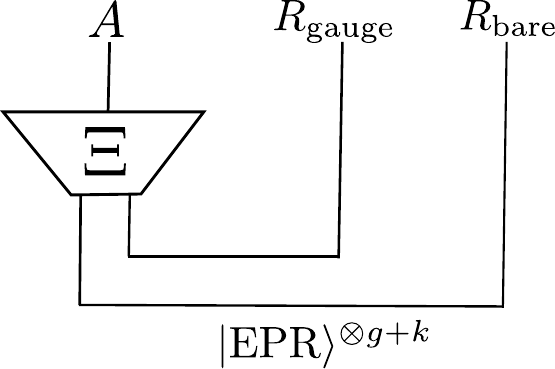}
\end{align}
where $|\Phi\rangle$ is an $(n+g+k)$-qubit state supported on the system $A$ and the references $R_{\mathrm{gauge}}$ and $R_{\mathrm{bare}}$.
Let $\mathcal{S}_{\mathrm{code}}$ be the original stabilizer group of the subsystem code. 
The stabilizer group $\mathcal{S}$ for the Choi state $|\Phi\rangle$ is 
\begin{align}
\mathcal{S} = \Big\langle \mathcal{S}_{\mathrm{code}}\otimes I \otimes I, 
\bar{P}_{\mathrm{gauge}} \otimes P_{\mathrm{gauge}} \otimes I ,
\bar{Q}_{\mathrm{bare}} \otimes I \otimes Q_{\mathrm{bare}}
\Big\rangle \label{eq:Choi_sub}
\end{align} 
where $\bar{P}_{\mathrm{gauge}}$ and $\bar{Q}_{\mathrm{bare}}$ are gauge and bare logical operators. 

\begin{lemma}
The encoded information in a subsystem code is retained after projective measurements if and only if 
\begin{align}
I(A,R_{\mathrm{bare}})=2k
\end{align}
in the post measurement Choi state $|\Phi'\rangle$.
\end{lemma}


Finally, we represent the above condition $I(A,R_{\mathrm{bare}})=2k$ in terms of stabilizer generators. 
For the updated stabilizer group $\mathcal{S}'$, consider its restrictions on $R_{\mathrm{gauge}}R_{\mathrm{bare}}$, $R_{\mathrm{gauge}}$, and $R_{\mathrm{bare}}$ respectively. Letting $\lambda_{R_{\mathrm{gauge}}R_{\mathrm{bare}}}$, $\lambda_{R_{\mathrm{gauge}}}$, and $\lambda_{R_{\mathrm{bare}}}$ be the number of independent generators for $\mathcal{S}'_{R_{\mathrm{gauge}}R_{\mathrm{bare}}}$, $\mathcal{S}'_{R_{\mathrm{gauge}}}$, and $\mathcal{S}'_{R_{\mathrm{bare}}}$ respectively, we have
\begin{align}
S_{R_{\mathrm{gauge}}R_{\mathrm{bare}}} =
g + k - \lambda_{R_{\mathrm{gauge}}R_{\mathrm{bare}}},\quad
S_{R_{\mathrm{gauge}}} =
g - \lambda_{R_{\mathrm{gauge}}}, \quad
S_{R_{\mathrm{bare}}} =
 k - \lambda_{R_{\mathrm{bare}}}.
\end{align}
One must have $S_{R_{\mathrm{bare}}}=k$ in order for $R_{\mathrm{bare}}$ to be maximally entangled with some other subsystem, and thus $\lambda_{R_{\mathrm{bare}}}=0$. 
Furthermore, since the post measurement Choi state is pure, we have 
\begin{align}
I(A,R_{\mathrm{bare}}) = S_{A} + S_{R_{\mathrm{bare}}} - S_{AR_{\mathrm{bare}}} 
=S_{R_{\mathrm{gauge}}R_{\mathrm{bare}}} + k  - S_{R_{\mathrm{gauge}}}
\end{align}
which implies 
$
S_{R_{\mathrm{gauge}}R_{\mathrm{bare}}}  - S_{R_{\mathrm{gauge}}} = k$.
Hence we arrive at the following condition 
\begin{align}
\lambda_{R_{\mathrm{gauge}}R_{\mathrm{bare}}} = \lambda_{R_{\mathrm{gauge}}}, \qquad \lambda_{R_{\mathrm{bare}}}=0.
\end{align}

\begin{lemma} \label{lemma:subsystem}
The encoded information in a subsystem code is retained in the post measurement state if and only if, in the Choi state, for any stabilizer operator supported on  $R_{\mathrm{gauge}}R_{\mathrm{bare}}$, it does not have non-trivial support on $R_{\mathrm{bare}}$.
\end{lemma}

\textbf{Information preservation condition:}
Finally, we state our main result concerning the information preservation condition in a subsystem code. 

\begin{theorem}\label{theorem:subsystem}
The following three statements, concerning Pauli measurements on a subsystem code, are equivalent: 
\begin{enumerate}[i)]
\item The measured Pauli operator group $\mathcal{M}$ does not contain any non-trivial dressed logical operator, namely
\begin{align}
\mathcal{M}\cap \mathcal{L}_{_{\mathrm{dressed}}} \subseteq \mathcal{G}.
\end{align}
\item For any non-trivial bare logical operator $\ell_{\mathrm{bare}}$, there always exists an equivalent bare logical operator $\ell_{\mathrm{bare}}' \sim \ell_{\mathrm{bare}}$ that commutes with $\mathcal{M}$, namely 
$ 
\ell_{\mathrm{bare}}'\in \mathcal{C}(\mathcal{M})$.
\item The subsystem code retains the encoded information after the measurements with respect to $\{M_{1},\cdots, M_m\}$, namely, in the Choi state, 
\begin{align}
I(A, R_{\mathrm{bare}}) = 2k
\end{align}
where $A$ is the system and $R_{\mathrm{bare}}$ is the reference for bare logical qubits.  
\end{enumerate}
\end{theorem}

The proof is presented in appendix~\ref{app:proof}. As a corollary, we obtain the cleaning lemma for measurements in a subsystem code.

\begin{lemma}
The following two statements are equivalent:
\begin{enumerate}[a)]
\item The subsystem code retains the encoded information after the projective measurements that acts on qubits on $A$ with respect to $\mathcal{M}$.  

\item For any non-trivial bare logical operator $\ell_{\text{bare}}$, 
there always exists an equivalent expression logical operator $\ell_{\text{bare}}'\sim \ell_{\text{bare}}$ which can be written as
\begin{align}
\ell'_{\text{bare}} = {\ell'_{\text{bare}}}_{A^c} \otimes {\ell'_{\text{bare}}}_{A}, \qquad \text{where} \quad {\ell'_{\text{bare}}}_{A^c}\not=I_{A^c}, \quad {\ell'_{\text{bare}}}_{A} \in \mathcal{M}.
\end{align}
Furthermore, ${\ell'_{\text{bare}}}_{A^c} \otimes I_{A}$ is a bare logical operator for the post measurement code. 
\end{enumerate}
\end{lemma}

\subsection{Measurement threshold v.s. erasure threshold}

In studies of QECCs, one is often interested in erasure errors.
Suppose that physical qubits are removed from the code randomly with some probability $p_{e}$. 
The central question is whether the remaining $(1-p_e)n$ qubits retains the information.
Namely, one is typically interested in the erasure threshold $p_{e}^{th}$ below which the encoded information is retained with high probability. 
Suppose that physical qubits in a subset $Q_e$ are removed from a QECC. The remaining system retains the encoded information if and only if $Q_e$ does not support any (dressed) logical operator~\cite{Yoshida2010, Haah2010}. 
Comparing this condition with that for the information preservation under measurement, we arrive at the following simple, but fundamental relation. 

\begin{theorem}
The measurement threshold, for both stabilizer and subsystem codes, is always larger than the erasure threshold, namely
\begin{eqnarray}
	p^{th}_e \leq p^{th}_m \ .
\end{eqnarray}
\end{theorem}

Note that the erasure threshold must satisfy $p^{th}_e\leq \frac{1}{2} $ due to the no-cloning theorem.
Several examples of QECCs considered in this paper achieve $p^{th}_e=\frac{1}{2}$ and $p^{th}_m = 1$ for some relative measurement frequencies $(\alpha_X,\alpha_Y,\alpha_Z)$.

\section{Randomly Monitored Concatenated Codes}\label{sec:concatenation}

In this section, we study the measurement threshold for several examples of concatenated quantum codes and find that the measurement threshold $p_{m}^{th}$ is significantly higher than the erasure threshold $p_e^{th}$. Namely, we find that both the five-qubit code and the seven-qubit code achieve $p_{m}^{th}=1$ under uniformly random Pauli measurements.

\subsection{Five-qubit code}

We begin by studying the concatenated five-qubit code. In particular, we will find that the measurement threshold is $p_m^{th}=1$ under random measurements for arbitrary relative measurement frequencies $(\alpha_X,\alpha_Y,\alpha_Z)$. 
This suggests that the logical information is preserved under random measurements as long as $p_m=p_X + p_Y + p_Z < 1$. 


\textbf{Measurement threshold:}
The stabilizer group and logical operators are
\begin{eqnarray}
	&&S = \langle X_1Z_2Z_3X_4I_5, \ I_1X_2Z_3Z_4X_5, X_1I_2X_3Z_4Z_5, \ Z_1X_2I_3X_4Z_5  \rangle \cr 
	&&\bar{X} = X_1X_2X_3X_4X_5, \ \bar{Y} = Y_1Y_2Y_3Y_4Y_5, \ \bar{Z} = Z_1Z_2Z_3Z_4Z_5.
\end{eqnarray}
Under a uniformly random probability distribution with $p_X=p_Y=p_Z=\frac{p_m}{3}$, we find the probabilities of measuring $X,Y,Z$ logical operators are also uniform. Namely, we obtain
\begin{eqnarray}
	p'_m = \frac{1}{9} (10 p_m^3 - p_m^5)
\end{eqnarray}
after one concatenation where  $(p_{\bar{X}},  p_{\bar{Y}}, p_{\bar{Z}})= (\frac{p'_m}{3},\frac{p'_m}{3},\frac{p'_m}{3})$.
This equation has two fixed points satisfying $p'_m=p_m$ at $p_m=0,1$ with $p_m=1$ being an unstable fixed point. 
Namely, whenever $p_m < 1$, the updated probability $p_m'$ decreases and eventually reaches to $0$ in the limit of many concatenations.
Hence, the measurement threshold of the five-qubit code is $p_m^{th}=1$ under uniformly random measurements. 

\begin{figure}[h!]
\centering
\includegraphics[width = 8cm]{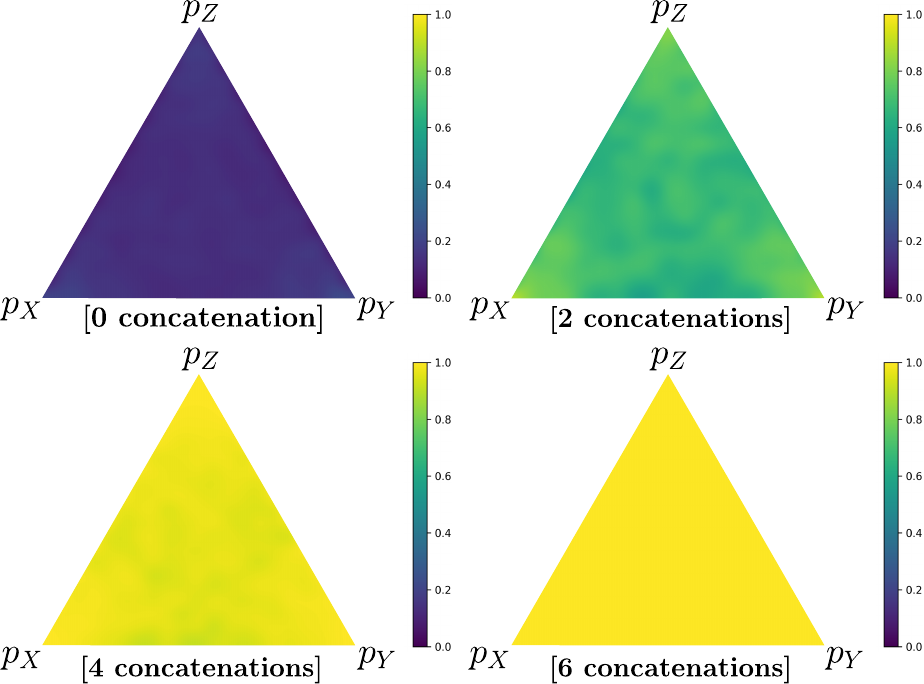}
\caption{Information preservation probability of the five-qubit code when $p_m = 0.95$. To obtain the data, we recursively computed the logical measurement probability
at each round of concatenation by performing random measurements, following the logical measurement
probability obtained in the preceding round. The random sampling size at each round of concatenations
was $10^3$.}
\label{fig:5qubit_info}
\end{figure}

In fact, we find that, for \emph{arbitrary} probability distributions $(p_X,p_Y,p_Z)$, the measurement threshold remains to be $p_m^{th}=1$. 
In other words, as long as $p_X + p_Y + p_Z < 1$, the probability of measuring a logical operator approaches 
$(p_{\bar{X}},  p_{\bar{Y}}, p_{\bar{Z}})\rightarrow (0,0,0)$ in the limit of many concatenations.
We have numerically confirmed this by computing the flow of the information preservation probabilities (Fig.~\ref{fig:5qubit_info}).
Specifically, we started from the initial probability with $p_m=0.95$ and computed the information preservation probability after concatenating the codes. 
After multiple times of concatenations, we find that the information preservation probability approaches to $1$ in all the probability distributions $(p_X,p_Y,p_Z)$. 
Hence, the concatenated five-qubit code has an extraordinary robustness against decoherence caused by measurements, satisfying $p_m^{th}=1$ for relative measurement frequencies $(\alpha_X,\alpha_Y,\alpha_Z)$. 
This is in contrast with that the erasure threshold of the five-qubit code is $p_e^{th}=\frac{1}{2}$. 

It is worth mentioning that a hyperbolic tiling of the five-qubit code is often used as a toy model of the AdS/CFT correspondence~\cite{Pastawski_2015}. 
For such a toy model of quantum error-correcting codes, previous works have found $p_m^{th}=1$ and $p_e^{th}=\frac{1}{2}$ via combination of analytical arguments and numerical evidences~\cite{Pastawski_2015, Antonini_2022}.

\textbf{Logical measurement probability:}
Even though the five-qubit code exhibits an extraordinary information preservation ability under random measurements, the encoded quantum information will be eventually measured when all physical qubits are measured. 
Here, we wish to study the probability distribution $(p_{\bar{X}},  p_{\bar{Y}}, p_{\bar{Z}})$ for the measured logical qubit when each physical qubit is measured with the probability distribution $(p_{X},  p_{Y}, p_{Z})$ satisfying $p_m=p_X+p_Y+p_Z=1$. 
We will find that, for all the distributions of $(p_{X},  p_{Y}, p_{Z})$ except for those with $p_X=0$, $p_Y=0$, and $p_Z=0$, the probability distribution of the logical qubit measurement approaches to a uniform distribution of $(p_{\bar{X}},  p_{\bar{Y}}, p_{\bar{Z}})= (\frac{1}{3},\frac{1}{3},\frac{1}{3})$ in the limit of many concatenations. 

\begin{figure}[h!]
\centering
\subfloat{\includegraphics[width = 5.5cm]{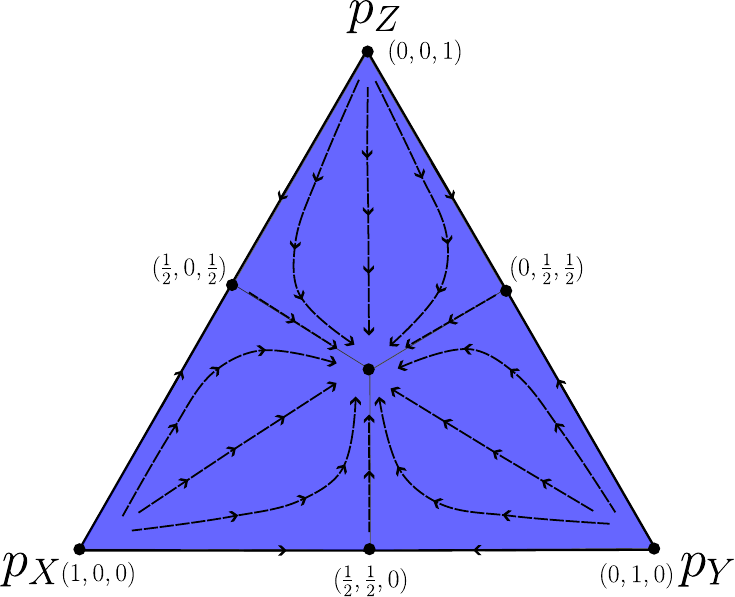}}
\qquad
\subfloat{\includegraphics[width = 6.0cm]{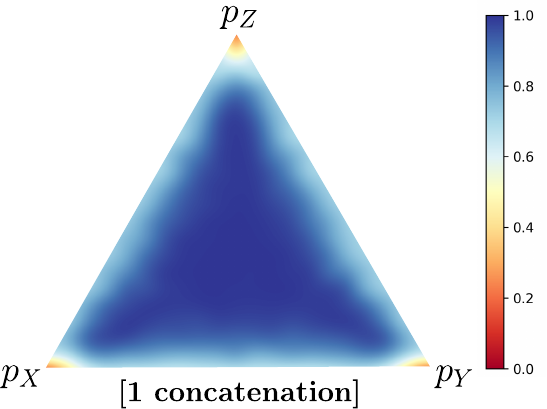}}
\caption{(Left) The flow of logical measurement probability of the concatenated five-qubit code. (Right) Uncertainty of measured logical operators after one concatenation of the code. Uncertainty of measured logical operators is quantified by Renyi-2 entropy, $S^{(2)}(p_{\bar{X}},p_{\bar{Y}},p_{\bar{Z}}) = - \log_3(p^2_{\bar{X}} + p^2_{\bar{Y}} + p^2_{\bar{Z}})$. The data were obtained by recursively computing logical measurement probability with random sampling size $10^3$ at each round of concatenations. }
\label{fig:5qubit}
\end{figure}

When all the qubits are measured in the same basis, the logical qubit is measured in one particular basis. Namely, for $(p_{X},  p_{Y}, p_{Z}) = (1,0,0), (0,1,0),(0,0,1)$, we find $(p_{\bar{X}},  p_{\bar{Y}}, p_{\bar{Z}})=(1,0,0),(0,1,0),(0,0,1)$. 
Next, when physical qubits are measured in a pair of randomly chosen Pauli basis, say in the $X$ and $Y$ basis, we find that the logical qubit will be measured in the two Pauli basis with equal probabilities. 
Namely, for $(p_{X},  p_{Y}, 0)$ with $p_{X},  p_{Y}\not =0$, we find that each round of concatenations generates a flow of the probability distributions toward the fixed point of $(p_{\bar{X}},  p_{\bar{Y}}, p_{\bar{Z}})=(\frac{1}{2},\frac{1}{2},0)$. 
Finally, for any other distribution of $(p_{X},  p_{Y}, p_{Z})$ with $p_X,p_Y,p_Z\not=0$, we find that the probability distribution flows to the uniform distribution $(p_{\bar{X}},  p_{\bar{Y}}, p_{\bar{Z}})= (\frac{1}{3},\frac{1}{3},\frac{1}{3})$. 
The flows of the probability distributions $(p_{X},  p_{Y}, p_{Z}) \rightarrow (p_{\bar{X}},  p_{\bar{Y}}, p_{\bar{Z}})$ under concatenations can be schematically visualized as in Fig.~\ref{fig:5qubit}.

\subsection{Seven-qubit code}

We also study the concatenated seven-qubit code, another example achieving $p_m^{th}=1$ under uniformly random measurements. 

\textbf{Measurement threshold:}
The stabilizer group and logical operators are
\begin{eqnarray}
	&&S = \langle X_1X_2X_3X_4, \ X_2X_3X_5X_6, \ X_3X_4X_6X_7,
	Z_1Z_2Z_3Z_4, \ Z_2Z_3Z_5Z_6, \ Z_3Z_4Z_6Z_7  \rangle \cr 
	&&\bar{X} = X_1X_2 \dotsb X_7, \bar{Y} = Y_1Y_2 \dotsb Y_7, \ \bar{Z} = Z_1Z_2 \dotsb Z_7.
\end{eqnarray}
The seven-qubit code can be viewed as the smallest 2D color code with a triangular boundary.

\begin{figure}[h!]
\centering
\includegraphics[width = 6cm]{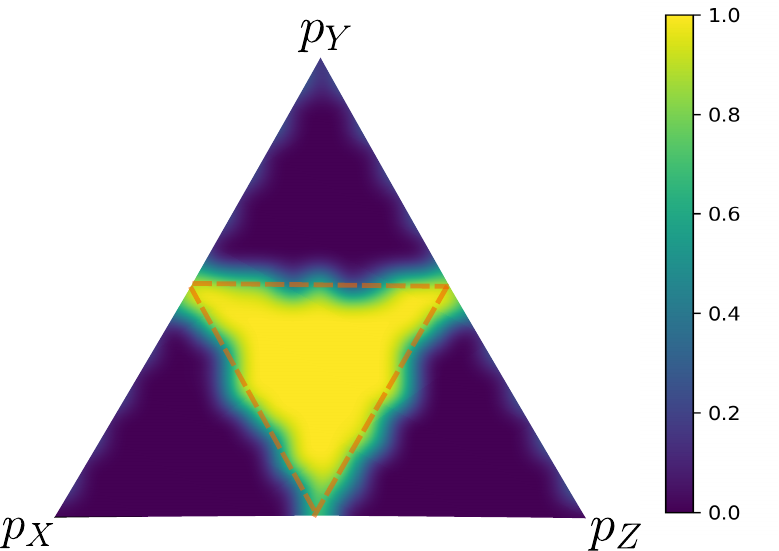 }
\caption{Information preservation probability of the seven-qubit code after seven concatenations when $p_m = 0.95$. 
The dashed lines indicate phase boundaries corresponding to $p_X = \frac{1}{2}$, $p_Y = \frac{1}{2}$, and $p_Z = \frac{1}{2}$. 
The data were obtained by recursively computing logical measurement probability with random sampling size $1.5 \times 10^3$ at each round of concatenations.}
\label{fig:7qubit_info}
\end{figure}

We have numerically computed the probability of measuring logical operators. 
We start with the initial probability distribution $(p_X,p_Y,p_Z)$ with $p_m = 0.95$ and obtained the logical probability distribution $(p_{\bar{X}},p_{\bar{Y}},p_{\bar{Z}})$ after seven rounds of concatenations. 
Fig.~\ref{fig:7qubit_info} shows the probability of retaining the logical qubit, suggesting that the information preservation condition is $p_X, p_Y, p_Z < \frac{1}{2}$, and $p_m < 1$.
Namely, the measurement threshold under uniformly random measurements is $p_m^{th}=1$.
This is in contract with that the erasure threshold of the seven-qubit code is $p_e^{th}=\frac{1}{2}$. 
In section \ref{sec:color code}, we will find that this condition qualitatively coincides with the one for the 2D color code. 

\begin{figure}[h!]
\centering
\subfloat{\includegraphics[width = 6cm]{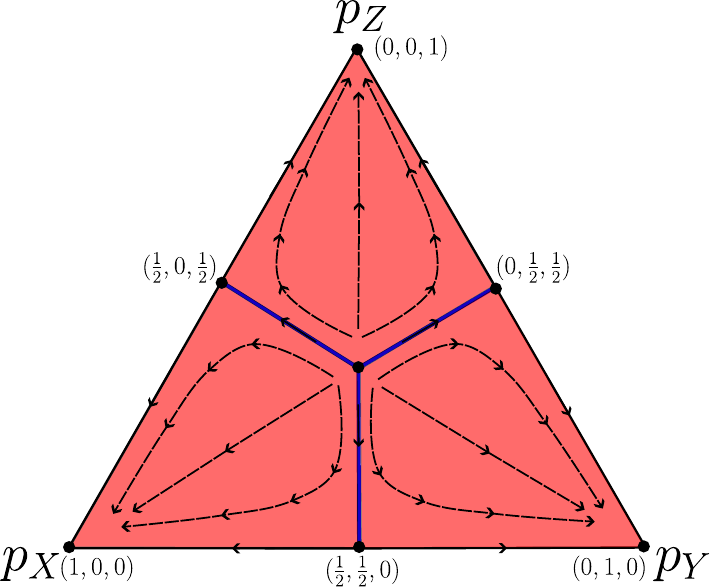}}
\qquad \qquad 
\subfloat{\includegraphics[width = 6.5cm]{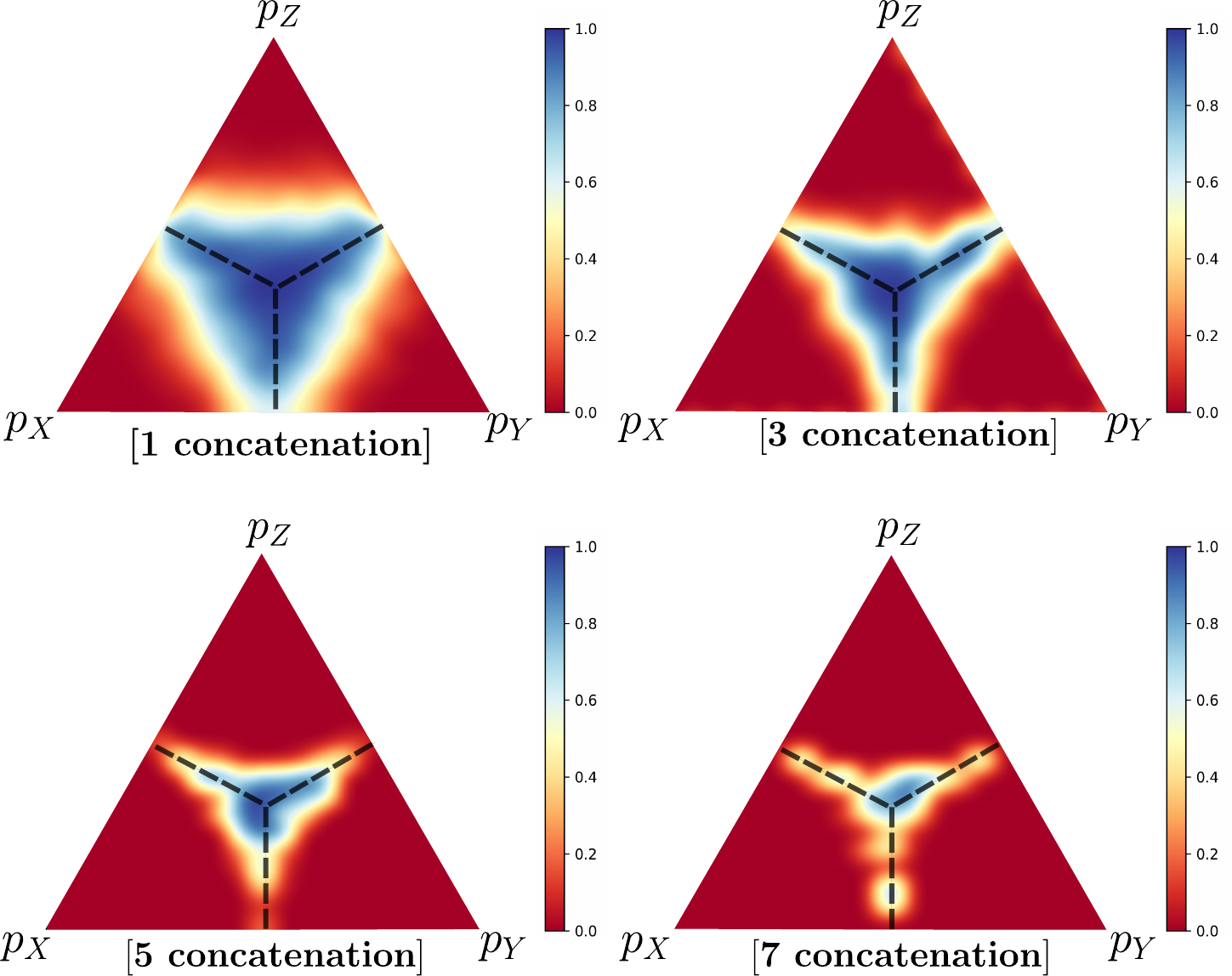}}
\caption{(Left) The flow of logical measurement probability of the concatenated seven-qubit code.
(Right)
The uncertainty of the basis of measured logical operators after concatenations, measured in terms of the R\`{e}nyi-$2$ entropy.
The data were obtained by recursively computing logical measurement probability with random sampling size $10^3$ at each round of concatenations.
}
\label{fig:7qubit}
\end{figure}

\textbf{Logical measurement probability:}
Next, we study the logical measurement probability distribution $(p_{\bar{X}},p_{\bar{Y}},p_{\bar{Z}})$ when all the physical qubits are measured with $(p_{X}, p_{Y}, p_{Z})$ at $p_m=1$ in the seven-qubit  code.
Unlike the five-qubit code case, we find that the uniform probability distribution $(\frac{1}{3},\frac{1}{3},\frac{1}{3})$ is an unstable fixed point in the flow generated by concatenations in the seven-qubit code. 
In fact, for generic distributions of $(p_{X},  p_{Y}, p_{Z})$, the probability distributions converge to $(1,0,0)$ , $(0,1,0)$, or $(0,0,1)$. 
The phase boundaries of measured logical operator are shown in Fig. \ref{fig:7qubit}.
Here it is useful to characterize the suppression of randomness with an entropic quantity. 
Specifically, we consider the R\'{e}nyi-$2$ entropy:
\begin{eqnarray}
S^{(2)}(p_{\bar{X}},p_{\bar{Y}},p_{\bar{Z}}) = - \log_3 (P_{\bar{X}}^2 + P_{\bar{Y}}^2 + P_{\bar{Z}}^2) 
\end{eqnarray}
where uniformly random distributions will give $S^{(2)}=1$.
Fig. \ref{fig:7qubit} shows how $S^{(2)}(p_{\bar{X}},p_{\bar{Y}},p_{\bar{Z}})$ evolves under the flow generated by concatenations.

\subsection{15-qubit code}

Finally, we study a concatenated 15-qubit Reed-Muller code. We find that this code has measurement threshold less than $1$ under uniformly random measurements.

\begin{figure}[h!]
\centering
\includegraphics[width = 4.5cm]{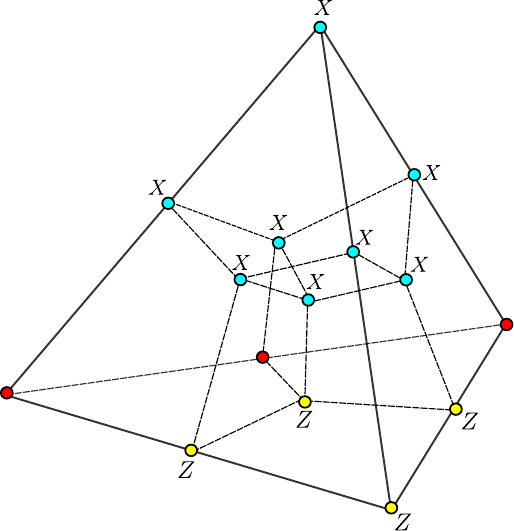}
\caption{The 15-qubit Reed-Muller code as the 3D color code. $X$ stabilizer generators act on each body cell (blue colored) and $Z$ stabilizer generators act on each face cell (yellow colored).
}
\label{15-qubit}
\end{figure}

\textbf{Measurement threshold:}
A convenient way to characterize the 15-qubit code is to view it as the 3D color code~\cite{Bombin_2007} supported on a tetrahedron as shown in Fig. \ref{15-qubit} where qubits live on vertices. 
The $X$ stabilizer generators are defined by weight-eight $X$ operators acting on each body cell, while $Z$ stabilizer generators are defined by weight-four $Z$ operators acting on each face cell. 

\begin{figure}[h!]
\centering
\subfloat{\includegraphics[width = 6.5cm]{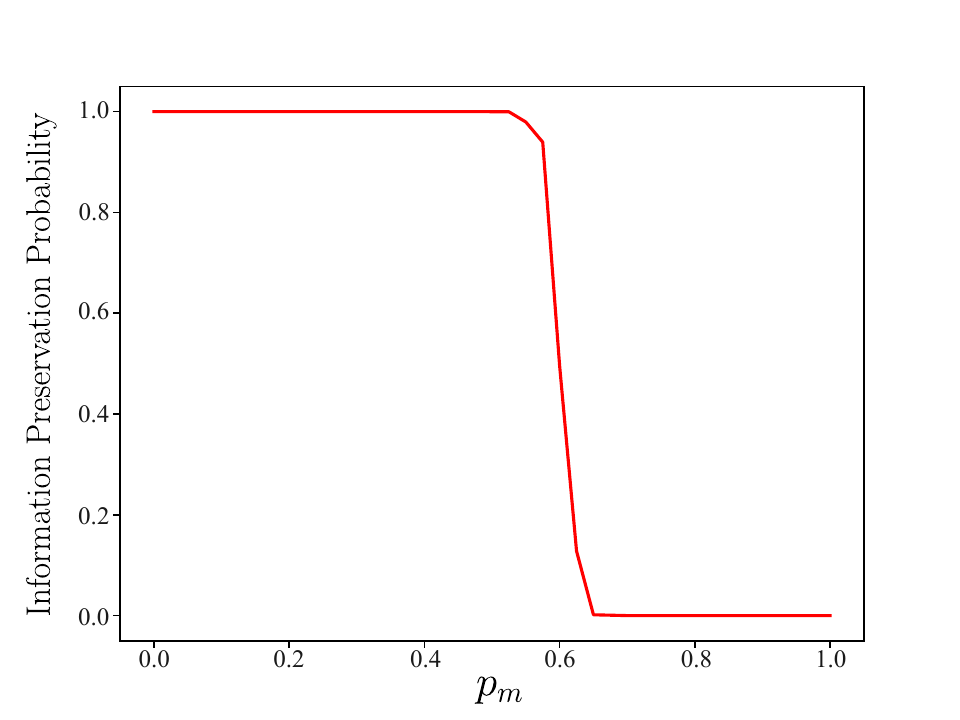}}
\subfloat{\includegraphics[width = 5.5cm]{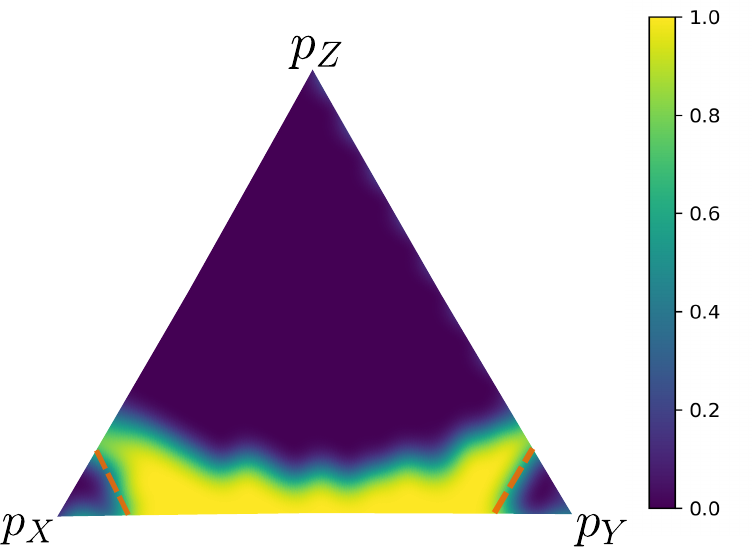}}
\caption{(Left) 
Information preservation probability of the 15-qubit code after three rounds of concatenations. 
(Right) Information preservation probability of the 15-qubit code after five rounds of concatenations when $p_m = 0.95$. 
The orange dashed lines correspond to $\ p_X= 0.8, \ p_Y= 0.8$. The data were obtained by recursively computing logical measurement probability with random sampling size $10^3$ at each round of concatenations. }
\label{fig:15data}
\end{figure}

One important observation is that the logical $Z$ operator has a smaller minimum weight compare to the ones for logical $X$ and $Y$ operators. 
Namely, the minimum weight of logical $Z$ operator is three while that of the logical $X$ and $Y$ operators is seven. Due to this asymmetric weights of different logical operators, we expect that the logical $Z$ operator will be more likely to be measured when uniformly random measurements are performed.
Namely, we expect that the measurement threshold $p_m^{th}$ will be lower than $1$ for uniform relative measurement frequencies $(\alpha_X, \alpha_Y, \alpha_Z)=(\frac{1}{3},\frac{1}{3},\frac{1}{3})$.
We numerically confirmed that this is indeed the case. 
Namely, we find $p_m^{th}\approx 0.6$, as shown in Fig.~\ref{fig:15data}.
We also numerically studied the measurement threshold $p_m^{th}$ for other relative measurement frequencies $(\alpha_X,\alpha_Y,\alpha_Z)$ after five rounds of concatenations.
We find that there is a region of relative measurement frequencies, with less frequent $Z$ measurements, which achieve $p_m^{th}=1$. 

\begin{figure}[h!]
\centering
\includegraphics[width = 10cm]{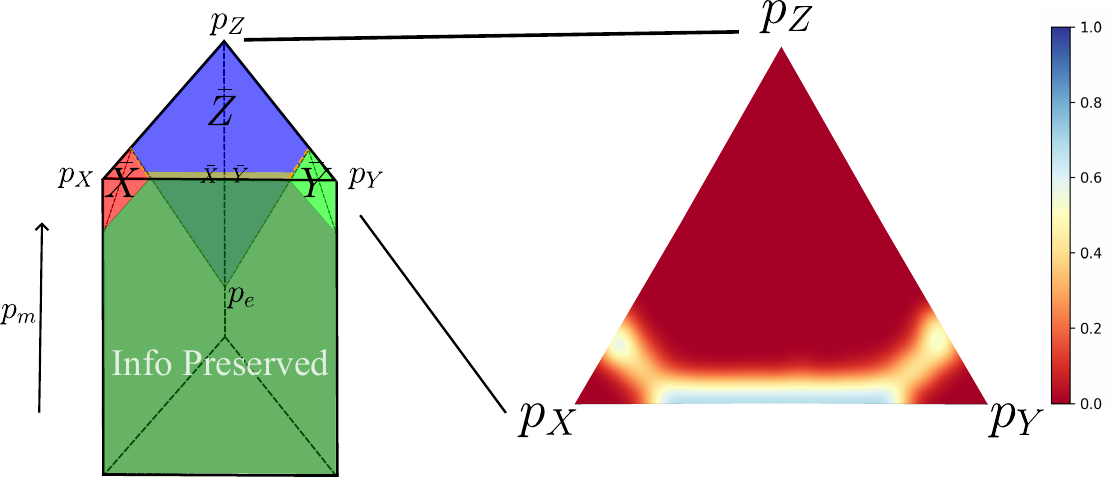}

\caption{Measured logical operators in the 15-qubit code after three rounds of concatenations. 
The numerical date suggests that some particular logical operator is measured except for a small region with $p_Z \lessapprox 0.95$,  $p_X, p_Y \lessapprox 0.8 $ (green brown region). 
The associated R\'{e}nyi-$2$ entropy $S^{(2)}(p_{\bar{X}},p_{\bar{Y}},p_{\bar{Z}})$ is also shown.
The data were obtained by recursively computing logical measurement probability with random sampling size $10^3$ at each round of concatenations. 
}
\label{fig:15_measured_op}
\end{figure}

\textbf{Logical measurement probability:}
We numerically studied the logical measurement probability distribution in the 15-qubit code (Fig.~\ref{fig:15_measured_op}). 
We find that there is a small region of measurement probabilities $(p_X,p_Y,p_Z)$ where both $\bar{X}$ and $\bar{Y}$ are measured with roughly equal probabilities. 
This region can be clearly seen by the  R\'{e}nyi-$2$ entropy.

\section{Randomly Monitored topological code}\label{sec:2D}

In this section, we turn our attentions to QECCs with geometrically local generators. 
Namely, we demonstrate that the 2D toric and color codes achieve $p_{m}^{th}=1$ under uniformly random measurements. 
We also discover an interesting correspondence between the randomness in measured logical probabilities $(p_{\bar{X}},p_{\bar{Y}},p_{\bar{Z}})$ and the measurement threshold $p_{m}^{th}$; we have $p_m^{th}=1$ under relative measurement frequencies $(\alpha_X,\alpha_Y,\alpha_Z)$ if and only if the measured logical probabilities $(p_{\bar{X}},p_{\bar{Y}},p_{\bar{Z}})$ have uncertainty at $p_m=1$. 
Finally, we study the 2D Bacon-Shor subsystem code and find $p_m^{th}=0$ for any relative measurement frequencies. 

\subsection{2D Toric Code}

We show that the 2D toric code, supported on a square lattice, achieves $p_m^{th}=1$ under uniformly random projective measurements. 
Furthermore, we determine the necessary and sufficient condition for retaining logical qubits in terms of $(p_X,p_Y,p_Z)$.

\begin{figure}[h!]
	\centering
	\includegraphics[width = 4.5cm]{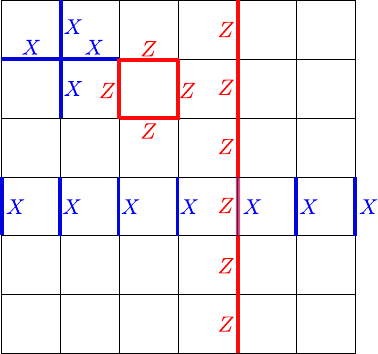}
	\caption{Stabilizer generators and logical operators of the 2D toric code.
 }
	\label{fig:toric}
\end{figure}

As shown in Fig.~\ref{fig:toric}, one physical qubit is assigned at each edge of the lattice in the toric code. Stabilizer generators are given by two different types of four-body Pauli operators: plaquette operators consisting of Pauli $Z$ are defined on each face, and star operators consisting of Pauli $X$ are defined on each vertex. Logical Pauli $Z$ operators are incontractible loops consisting of Pauli $Z$. Similarly, logical Pauli $X$ operators are incontractible loops of Pauli $X$ on the dual lattice.

\begin{claim}
 Suppose random single-qubit Pauli measurements are performed on the 2D toric code on a square lattice with probability $(p_X,p_Y,p_Z)$. The encoded information is preserved if and only if 
	\begin{eqnarray}
		p_X < \frac{1}{2}, \ p_Z < \frac{1}{2}, \text{ and }\  p_m < 1 .
	\end{eqnarray}
\end{claim}

We will first show that, when Pauli $X$ and $Z$ measurements are randomly performed on the toric code with probabilities $p_X, p_Z < \frac{1}{2}$, the unmeasured qubits still retain the logical qubits in the toric code defined on a deformed 2D lattice. 
We then argue that Pauli $Y$ measurements with $p_Y < 1$ on the 2D toric code do not destroy logical qubits. 

\begin{figure}[h!]
    \centering
    \includegraphics[width = 14cm]{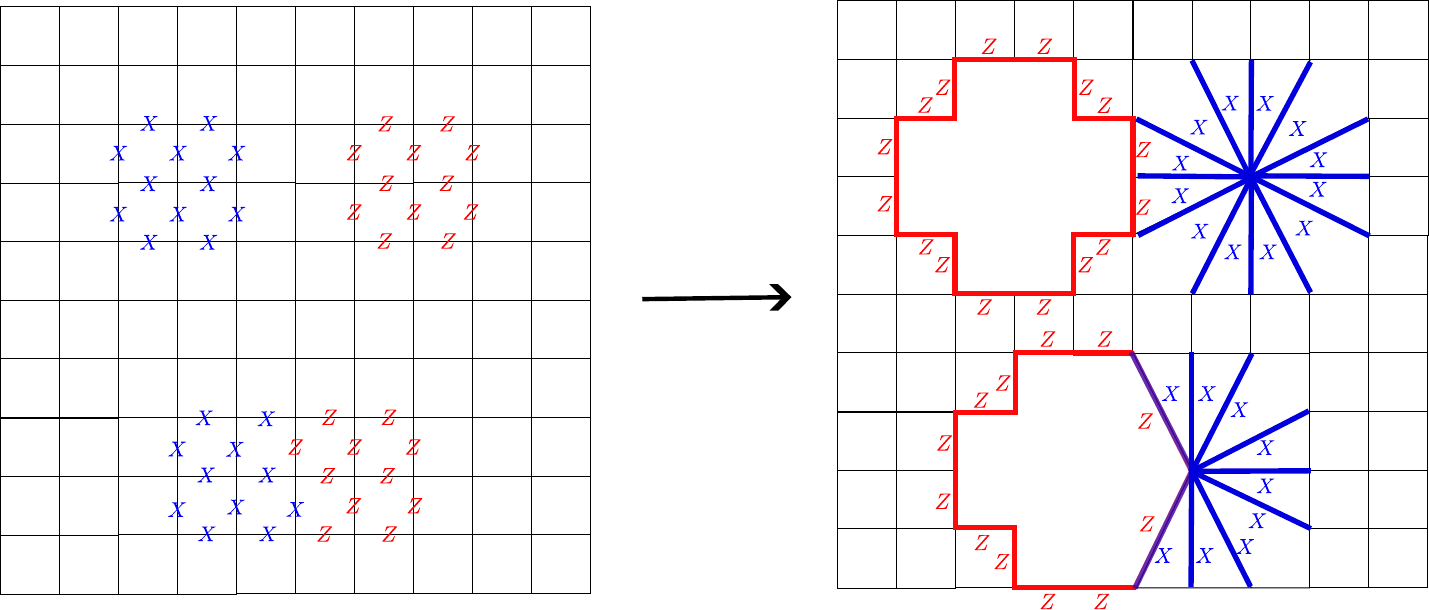}
    \caption{Lattice deformation induced by Pauli $X$ and $Z$ measurements}
    \label{fig:deformation}
\end{figure}

\textbf{$X$ and $Z$ measurements:}
When a single $X$ measurement is performed on an edge $e$, a pair of plaquette $Z$ stabilizers, containing this edge, do not commute with the measured $X_e$ operator, and thus are removed from the stabilizer group. 
Instead, one adds the product of two plaquette $Z$ stabilizer generators to form the updated stabilizer group. 
This effect can be visualized as follows
\begin{align}
\figbox{0.8}{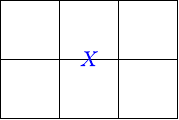} \longrightarrow \figbox{0.8}{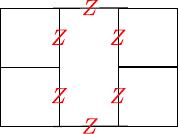} 
\end{align}
where the updated system is still the 2D toric code, but defined on a graph with an edge $e$ removed. 
Similarly, when a single $Z$ measurement is performed on an edge $e$, one removes a pair of star $X$ operators and adds their product to the stabilizer group. This leads to
\begin{align}
\figbox{0.8}{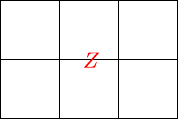} \longrightarrow \figbox{0.8}{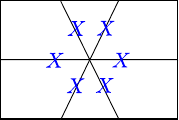}\ ,
\end{align}
where the edge $e$ is removed and two vertices on $e$ are merged into a single vertex. 
An example of such lattice deformations is shown in Fig.~\ref{fig:deformation}. 
It is worth noting that connected clusters of $X$ and $Z$ measurements create the so-called smooth and rough boundaries of the Toric code respectively, where large stabilizers around clusters can be interpreted as operators that measure the anyon types associated with the boundaries~\cite{Bravyi98}.

Recall that the bond percolation threshold of 2D square lattice is $\frac{1}{2}$.
Namely, if Pauli $X$ measurements are performed with a probability less than $\frac{1}{2}$, measured qubits will form a set of disjoint clusters where the probability of finding a large cluster is exponentially suppressed with respect to its linear size.
Namely, most of the clusters have finite sizes while there can be a small number of $\sim \log L$ size clusters. 
Hence, after the $X$ measurements, one will have a punctured 2D lattice with mostly finite size holes.  
Similarly, Pauli $Z$ measurements with a probability less than $\frac{1}{2}$ will create a punctured lattice with holes with identified vertices at the centers.
The upshot is that, when $p_X,p_Z<\frac{1}{2}$ (and $p_Y=0$), the remaining unmeasured qubits will form the toric code on a deformed lattice that has the geometry of a torus. 

\textbf{$Y$ measurements:}
Next, we analyze the probability of measuring a logical operator by $Y$ measurements while assuming $p_X=p_Z=0$. 
The key observation is that $Y$-type logical operators, consisting only of Pauli $Y$ operators, are very rare compared to $X$-type and $Z$-type logical operators. 
As such, the probability of measuring a $Y$-type logical operator remains exponentially suppressed even when $p_Y$ is close to $1$. 

\begin{figure}[h!]
	\centering
	\includegraphics[width =7.5cm]{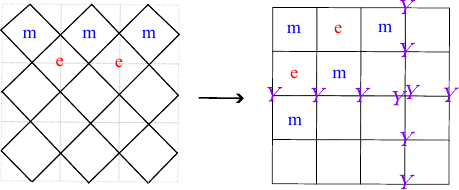}
	\caption{Stabilizer generators in the $45$ degree rotated toric code. Physical qubits are located at the each vertex of the lattice. The $Y$-commutant operators form straight lines, and are not deformable.}
	\label{fig:45toric}
\end{figure}

We begin by finding all the $Y$-type operators that commute with stabilizer generators. 
Such operators will be referred to as \emph{$Y$-commutant operators}.
Note that a $Y$-commutant operator can be either a stabilizer or a logical operator.
It will be convenient to consider a new lattice where qubits live on vertices as shown in Fig.~\ref{fig:45toric}. 
Note that the figure is $45$ degree rotated for simplicity of presentation. 
In this rotated lattice, both $X$-type and $Z$-type stabilizer generators are defined on faces and placed in an alternating pattern while physical qubits live on vertices.  
By inspecting commutation relations with stabilizer generators, one finds that $Y$-commutant operators can be generated by straight line-like operators, associated to each raw and column of the lattice, as shown in Fig.~\ref{fig:45toric}. 
Namely, one can consider $2L$ of such line-like operators where $L$ is the linear length of the rotated lattice. 
Note that only $2L-1$ of them are independent as a product of all the line-like $Y$-type operators is an identity.  
It is worth emphasizing that, unlike $X$- and $Z$-type logical operators, these line-like $Y$-commutant operators are not deformable. 

Suppose that a $Y$-commutant operator $\ell$ is constructed by using $a$ row operators and $b$ column operators ($0\leq a,b\leq L$). 
The weight of such an operator is
\begin{align}
	W(a,b) = (L-a)b + (L-b)a .
\end{align}
The probability of measuring a logical qubit can be bounded by adding probabilities of measuring individual $Y$-commutant operators. 
Namely, we have
\begin{eqnarray}
P_{\text{destroy}} &\leq & \sum_{0\leq a, b \leq L}  \binom La \binom Lb p^{W(a,b)} \cr 
&=& 2\sum_{0 \leq a,b \leq L/2}  \binom La \binom Lb p^{W(a,b)}
+  2\sum_{0 \leq a \leq L/2} \sum_{L/2 < b \leq L} \binom La \binom Lb p^{W(a,b)}.
\end{eqnarray} 
Using the ineqaulity $\binom nk < \left( \frac{ne}{k} \right)^k$, we can bound the first term as
\begin{eqnarray}
\sum_{0 \leq a, b\leq L/2}  \binom La \binom Lb p^{W(a,b)}
&<& \sum_{0 \leq a,b\leq L/2}  
\left( \frac{Le}{a} \right)^a \left( \frac{Le}{b} \right)^b p^{(L-a)b + (L-b)a} 
\cr
&<& \sum_{0 \leq a,b\leq L/2} \left( \frac{Le}{a} \right)^a \left( \frac{Le}{b} \right)^b p^{L(a+b)/2} \cr 
&=& \sum_{0 \leq a,b\leq L/2}  \left( \frac{Le}{a} p^{L/2} \right)^a \left( \frac{Le}{b} p^{L/2} \right)^b \ \rightarrow \ 0 \quad (L\rightarrow \infty)
\end{eqnarray}
where we used the fact that $\frac{Le}{a} p^{L/2} \approx p^{L/2}$ approaches zero exponentially. 
The second term can be bounded in a similar fashion. 
Hence, when $p_Y < 1$, the probability of measuring a $Y$-type logical operator is suppressed exponentially with respect to $L$. 

\begin{figure}[h!]
\centering
\includegraphics[width = 9cm]{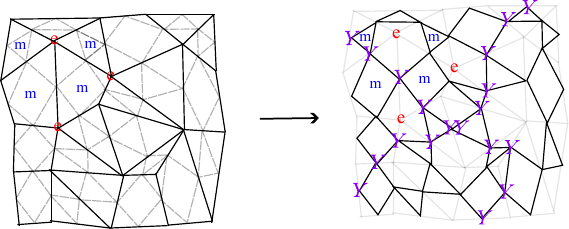}
\caption{The toric code on the deformed lattice. 
The $Y$-commutant operators form locally straight lines. }
\label{fig:deformed_lattice}
\end{figure}

\textbf{$X,Y,Z$ measurements:}
Finally, we consider the case of $p_X,p_Y, p_Z \not=0$. 
Here, we provide a heuristic argument to bound $P_{\text{destroy}}$ while providing numerical evidences later. 
After performing $X,Z$ measurements, we will have the 2D toric code supported on a deformed square lattice which has mostly finite densities in terms of the discretized metric.
To find $Y$ logical operators, we start by constructing a new lattice where qubits live on vertices as shown in Fig.~\ref{fig:deformed_lattice}. 
This new lattice has a structure locally similar to a square lattice where plaquette and star ($m$ and $e$) operators are placed in an alternating pattern. 
Most importantly, vertices are always $4$-valent. 
Due to this special structure, generators of $Y$-commutant operators can be constructed by line-like operators that always move straight at each vertex as shown in Fig.~\ref{fig:deformed_lattice}. 

In principle, one can construct all the $Y$-commutant generators by drawing a path of a straight line until it returns to the starting vertex from the opposite direction. 
Unlike the original square lattice, however, a straight line moving on a deformed lattice may form a closed loop of finite size that does not go around the torus. 
The probability of measuring such small $Y$ operators remains finite, and thus, our strategy to bound $P_{\text{destroy}}$ by adding all the measurement probabilities would not work.
Fortunately, such small $Y$ operators must be stabilizer operators since a logical operator in the deformed toric code must go around the torus. 
This suggests that, to bound $P_{\text{destroy}}$, we only need to consider probabilities of measuring $Y$-commutant operators whose sizes are at least $O(L)$. 

\begin{figure}[h!]
	\centering
	\includegraphics[width = 6cm]{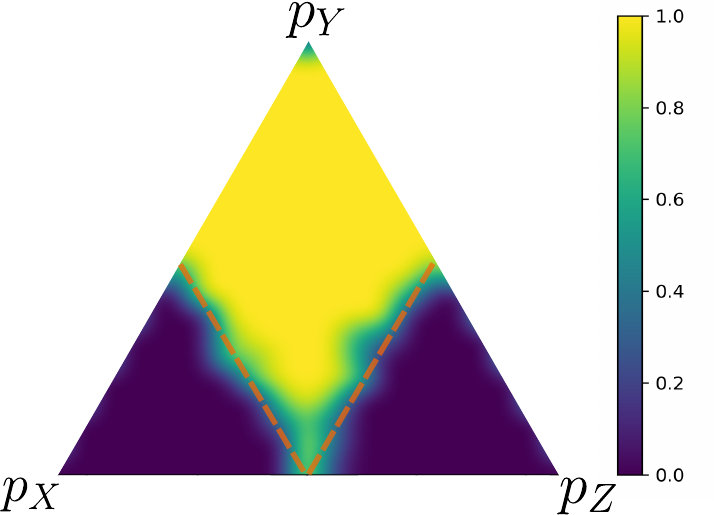}
    \caption{Information preservation probability in the 2D toric code when $p_m=0.95$. 
    The dashed lines correspond to the percolation threshold at $p_X = \frac{1}{2}$ and $p_Z = \frac{1}{2}$. Numerical simulation was performed on a $35 \times 35$ toric code with random sampling size $10^2$. }
	\label{fig:toric_ter}
\end{figure}

In constructing $Y$-commutant generators, the same edge can be used only once. 
Hence, there are at most $O(L)$ $Y$-generators. 
Here, we need to find the probability distributions of the weights of $Y$-commutant operators. 
We expect that, due to the lattice deformation, the number of large $Y$ generators is actually small, possibly $\sim L^{\alpha}$ with $\alpha <1$, since closed loops will be harder to form. 
We also expect that non-trivial $Y$ generators tend to have larger weights, possibly $\sim L^{\beta}$ with $\beta >1$. 
As such, measuring $Y$ logical operators will be even harder on a deformed lattice. 
Based on these observations, we claim that the 2D toric code on a randomly deformed lattice retains logical qubits after $Y$ measurements.  
We numerically confirmed this as shown in Fig.~\ref{fig:toric_ter}.

\begin{figure}[h!]
	\centering 
\subfloat{\includegraphics[width = 10cm]{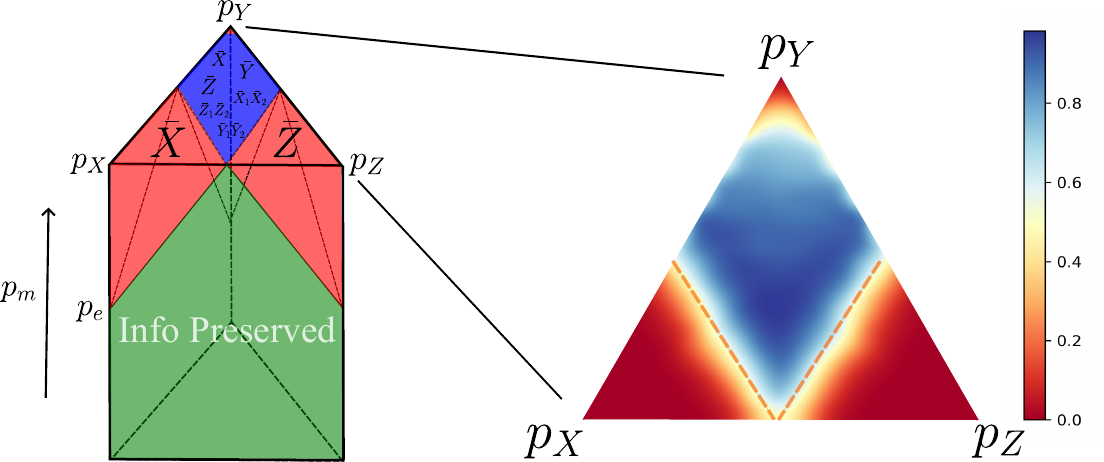}}
\subfloat{\includegraphics[width = 7cm, height = 4.5cm]{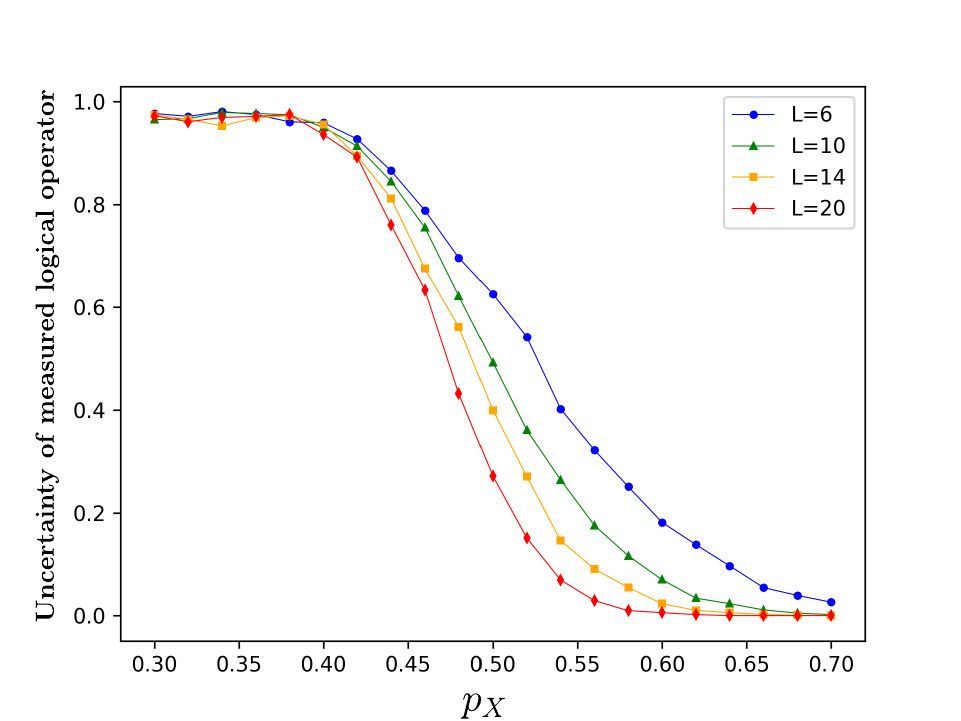}}
\caption{
(Left) 
A schematic picture showing the information preservation diagram of the 2D toric code under $(p_X,p_Y,p_Z)$ random measurements. 
Red: encoded information is measured by some particular logical operators. Blue: encoded information is measured with uncertainty in logical measurement probabilities. Green: encoded information is preserved. 
(Middle) The uncertainty of measured logical operators at $p_m = 1$ for a $10 \times 10$ toric code. The data was obtained using random sampling with sampling size $10^3$. 
(Right) The uncertainty transition at the $p_Y = p_Z$ line.
}
\label{fig:toricop}
\end{figure}

\textbf{Logical measurement probability:}
We also studied the logical measurement probability at $p_m=1$. 
The result is summarized in Fig.~\ref{fig:toricop}. 
We find that, for $p_X > \frac{1}{2}$ and $p_Z > \frac{1}{2}$, particular logical operators, namely $\bar{X}_1, \bar{X}_2$ and $\bar{Z}_1,\bar{Z}_2$, are measured respectively.
Also, for $p_Y \approx 1$, we find that $\bar{Y}_1, \bar{Y}_2$ are always measured at the large lattice size limit for odd $L$, and $\bar{X}_1 \bar{X}_2, \bar{Z}_1 \bar{Z}_{2}$ for even $L$. 
For all other distributions of $(p_X,p_Y,p_Z)$, i.e. $p_X < \frac{1}{2}$, $p_Z < \frac{1}{2}$, and $p_Y<1$, we observed that measured logical operators are uncertain. 
Specifically, we find that there are ten possible sets of logical operators to be measured; 
$\{\bar{P}_{1},\bar{Q}_{2}\}$ and
$\{\bar{X}_1\bar{X}_2, \bar{Z}_1\bar{Z}_2\}$ where $P,Q=X,Y,Z$.
The uncertainty is quantified by the R\'{e}nyi-$2$ entropy 
$S^{(2)} = - \log_4 (p_{\bar{X}_1}^2 + p_{\bar{Y}_1}^2 + p_{\bar{Z}_1}^2 + p_{\bar{X}_1 \bar{X}_2}^2)$, as shown in the heatmap in Fig.~\ref{fig:toricop}.
Here $p_{\bar{P}_1}$ represents the probability of measuring $\bar{P}_{1}$, combining three cases $\{\bar{P}_1, \bar{X}_2\},\{\bar{P}_1, \bar{Y}_2\}, \{\bar{P}_1, \bar{Z}_2\}$.

We have also studied the transition between the uncertain region and the no-uncertainty region by changing $p_X$ while fixing $p_Y=p_Z$. 
The R\'{e}nyi-$2$ entropy $S^{(2)}$ is plotted in Fig.~\ref{fig:toricop} where we observe a transition at $p_X\approx \frac{1}{2}$ as expected. 
We were not able to determine from the data if this transition is a smooth one or not, and we leave this question for future studies

\textbf{Information preservation diagram:}
We have addressed the information preservation probability and the logical measurement probability. 
Our result hints an intriguing relation between them, as captured in Fig.~\ref{fig:toricop}. 
In this diagram, we aim to characterize the fate of encoded information for arbitrary distributions $(p_X,p_Y,p_Z)$ including those with $p_m<1$ and $p_m=1$ as a 3D figure. 
The red shaded regions correspond to random measurement probability distributions where encoded information is measured by some particular logical operators. 
This includes two tetrahedron regions with $p_X>\frac{1}{2}$ and $p_Z>\frac{1}{2}$, and a single point corresponding to $p_Y=1$.
The green shaded region corresponds to $(p_X,p_Y,p_Z)$ where encoded information is still preserved with $p_X<\frac{1}{2}$, $p_Z<\frac{1}{2}$, and $p_m<1$. 
Finally, the blue shaded region emerges at the top boundary of the green region, corresponding to $(p_X,p_Y,p_Z)$ with the uncertainty in logical measurement probabilities. 
From this figure, one may speculate that there is some intrinsic relation between the measurement threshold $p_m^{th}$ and the uncertainty $S^{(2)}$ of measured logical operators. 
Namely, for relative measurement frequencies $(\alpha_X,\alpha_Y,\alpha_Z)$ that achieve $p_m^{th}=1$, we find $S^{(2)}>0$ when $p_m=1$. 
On the other hand, when $p_m^{th}<1$, we find $S^{(2)}=0$ when $p_m=1$. 
We will find a similar relation in the 2D color code. 

\subsection{2D Color Code}

\begin{figure}[h!]
\centering
\subfloat{\includegraphics[width = 5cm]{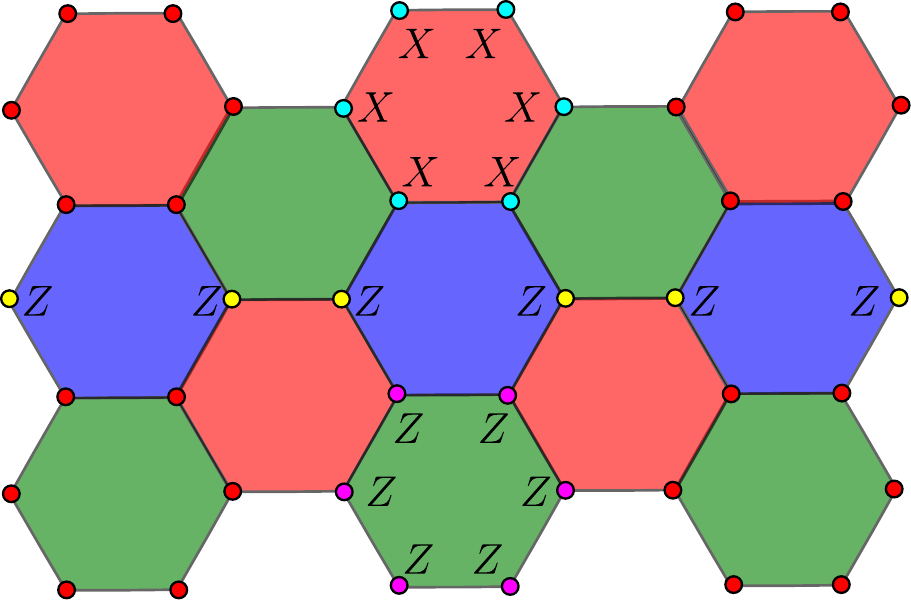}}
\qquad \qquad
\subfloat{\includegraphics[width = 5cm]{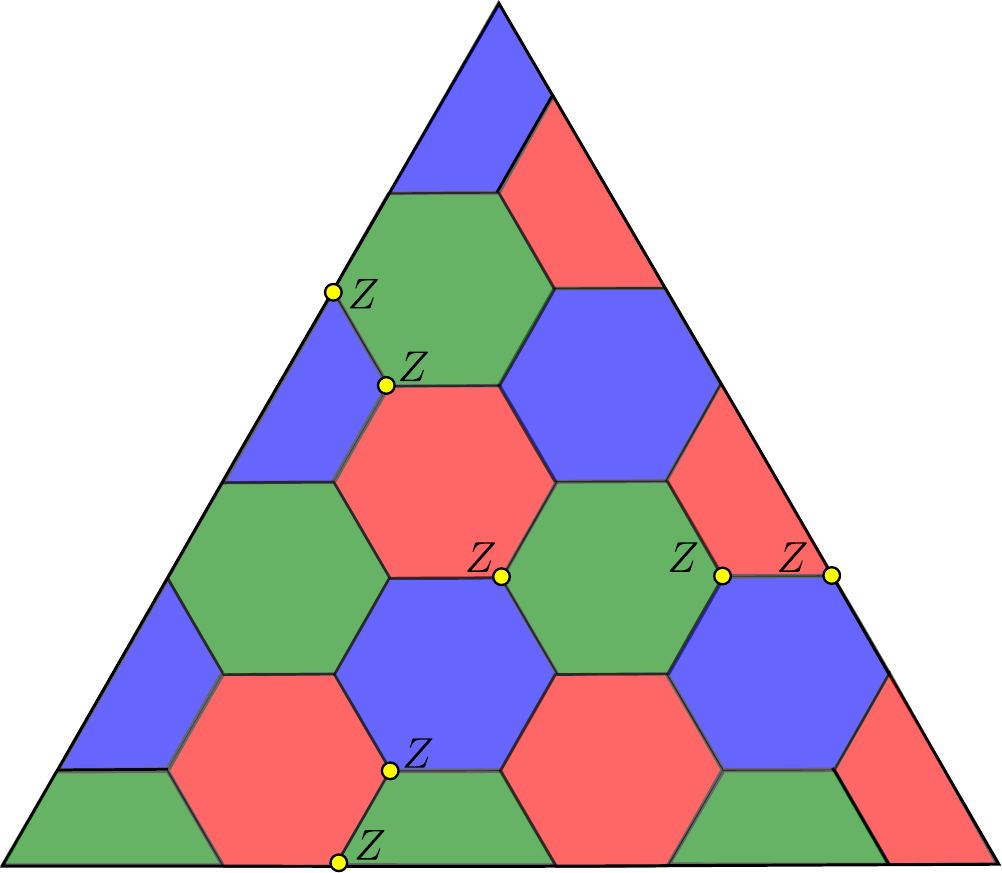}}
\caption{(Left) A topological color code defined on the 2D hexagonal lattice. (Right) A topological color code with triangular boundaries.   }
\label{fig:color}
\end{figure}

Next, we study an example of 2D topological codes that has symmetry under exchanges of $X,Y,Z$ operators. 
The 2D color code is defined on a graph satisfying valence and colorability conditions. Specifically, each vertex of a graph should belong to three edges, and each face should be three-colorable in a way that two adjacent faces do not share the same color (Fig.~\ref{fig:color}). A hexagonal lattice is one example that satisfies these conditions. Given such a graph, physical qubits reside at each vertex and stabilizer generators are defined on each face, consisting of only Pauli $Z$ or Pauli $X$. Logical operators are given by string operators that connects faces with the same colors. 

\begin{figure}[h!]
	\centering
	\includegraphics[width = 6cm]{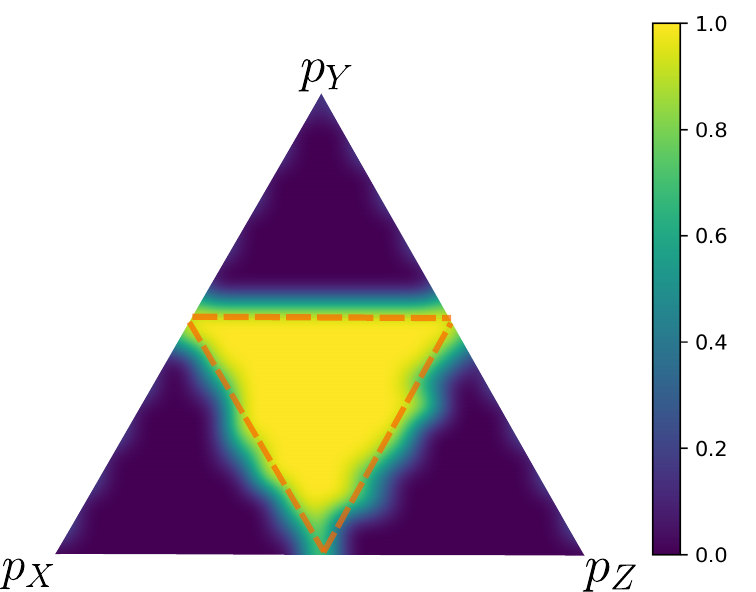}
	\caption{Information preservation diagram of 2D color code at $p_m=0.95$. Simulation was performed using a color code with triangular boundaries of length 35. The orange dashed lines correspond to $p_X = \frac{1}{2}, \  p_Y = \frac{1}{2}, \ p_Z = \frac{1}{2}$ lines. The data were obtained by random sampling with sampling size $10^2$.}
	\label{fig:color_info_ter}
\end{figure}

We numerically studied the information preservation condition for the 2D color code. 
Our result, summarized in Fig.~\ref{fig:color_info_ter}, suggests that encoded information will be preserved if and only if the random local measurements are performed with probability $p_X, p_Y, p_Z \lessapprox \frac{1}{2}$ and $p_m < 1$.

\begin{figure}[h!]
	\centering
\subfloat{\includegraphics[width = 9cm]{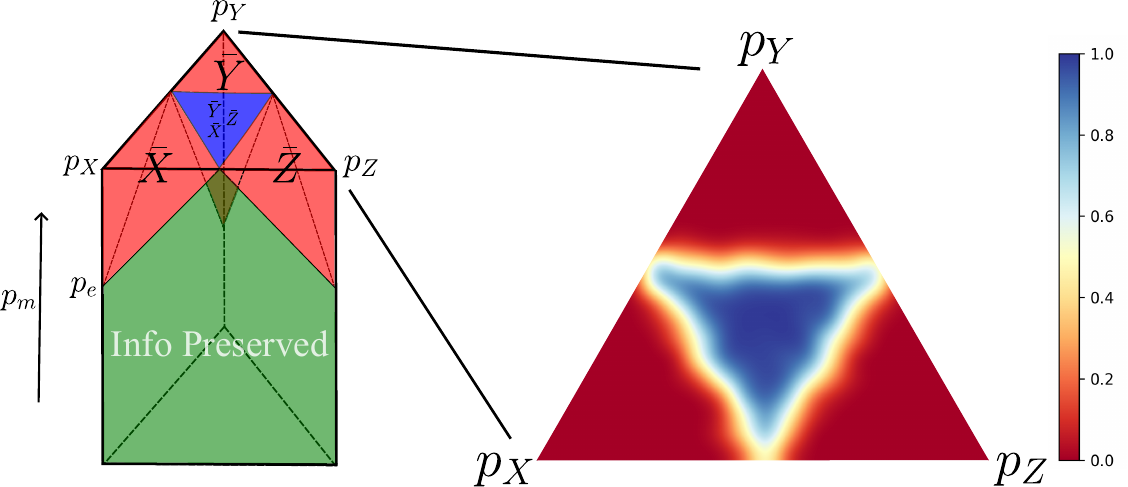}}
\subfloat{\includegraphics[width = 7cm, height = 4.5cm]{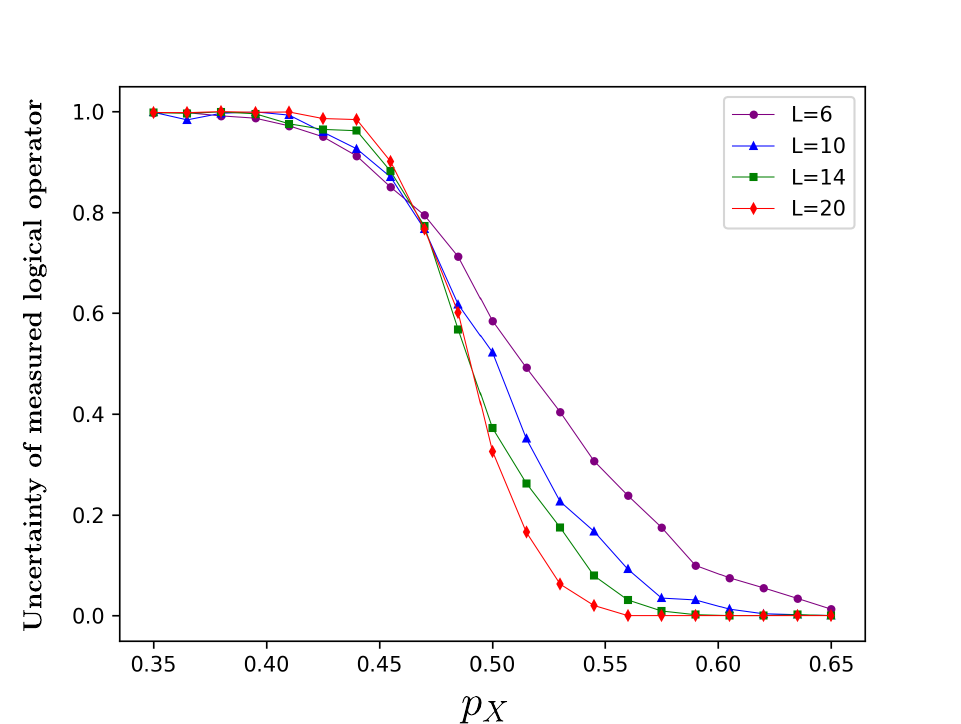}}
\caption{(Left) Information preservation diagram of the 2D color code. (Middle) Uncertainty of a measured logical operator at $p_m = 1$. The uncertainty of measured logical information is quantified by $S = - \log (p_{\bar{X}}^2 + p_{\bar{Y}}^2 + p_{\bar{Z}}^2)$. (Right) The uncertainty transition at $p_Y = p_Z$ line. } 
\label{fig:colorop}
\end{figure}

We also numerically studied the uncertainty of measured logical operator at $p_m = 1$. The result is summarized in Fig.~\ref{fig:colorop}. 
We find that, for $p_X, p_Y, p_Z \gtrapprox \frac{1}{2}$, encoded information will be measured in some particular logical operators. 
On the other hand, for $p_X, p_Y, p_Z \lessapprox \frac{1}{2}$, the logical measurement probabilities have some uncertainty, and encoded information will be measured by $\bar{X},\bar{Y},\bar{Z}$ with roughly equal probabilities.  
The associated R\'{e}nyi-$2$ entropy is plotted in Fig.~\ref{fig:colorop}. 

From Fig.~\ref{fig:color_info_ter} and Fig.~\ref{fig:colorop}, we again observe that relative measurement frequencies $(\alpha_X,\alpha_Y,\alpha_Z)$ achieving $p_m^{th}=1$ match to those possessing the uncertainty in $(p_{\bar{X}},p_{\bar{Y}},p_{\bar{Z}})$ at $p_m=1$.

\label{sec:color code}

\subsection{2D Bacon-Shor code}

\begin{figure}
\centering
\includegraphics[width= 3.5cm]{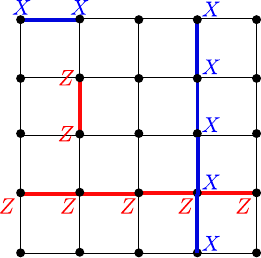}
\caption{Gauge operators and bare logical operators of 2D Bacon-Shor code.}
\label{fig:bacon}
\end{figure}

Finally, we discuss the measurement threshold of the 2D Bacon-Shor code. The gauge group is given by
\begin{align}
\mathcal{G} = \langle X_{i,j}X_{i+1,j}, Z_{i,j}Z_{i,j+1}   \rangle
\end{align}
as depicted in Fig.~\ref{fig:bacon}.
Here, we show that $p_{m}^{th}=0$. 

We think of performing $X$ measurement with probability $p_{X}>0$.
Observe that a tensor product of $X$ operators in the vertical direction commutes with all the gauge operators, and thus is a bare logical operator as shown in Fig.~\ref{fig:bacon}. 
Deforming this bare logical operator by $X$-type gauge operators can create a dressed logical operator that has single $X$ support at arbitrary location on each row. 
In other words, having an $X$ operator on each row creates a dressed logical operator. 
Randomly selecting qubits with probability $p_X$ will pick at least one qubit from each row with high probability. 
Hence, the measurement group $\mathcal{M}$ will contain a dressed logical operator, suggesting the loss of the encoded information for any $p_X>0$ and $p_Z>0$. 

Similar arguments can be applied to the case with $Y$ measurements. One can create a $Y$-type dressed logical operator by multiplying $X$-type and $Z$-type dressed logical operators that are supported on the same set of qubits. 
Specifically, a $Y$-type dressed logical operator has single $Y$ support on each row and column. 
When $p_Y>0$, out of $L$ qubits in each row, $\sim p_Y L$ qubits will be measured in $Y$.
For large $L$, the probability of having a row with no $Y$ measurements is exponentially suppressed. 
A similar observation applies to measurements in each column. 
Hence, the encoded information will be lost for any $p_Y>0$. 

\section{Haar random code and measurement}\label{sec:Haar}

In this section, we study the information preservation in QECCs when Haar randomness is involved in encoding or measurement. 
Specifically, we study two scenarios. 
The first problem considers a Haar random encoding of quantum information and studies the effect of projective measurements. 
In this case, we find that the encoded information is retained even when $n-(k+\epsilon)$ qubits are projectively measured. 
The second problem considers an arbitrary QECC (not necessarily Haar random code) where a part of the system is measured in globally Haar random basis, instead of locally Haar random basis. 
For this problem, we will derive a simple condition for the original information to be recoverable after projective measurements. 

For these problems, similar results have been already obtained in the literature. 
Here we would like to highlight conceptually relevant works, often referred to as environment-assisted quantum channels, on these problems, see~\cite{PhysRevA.72.052317} for instance.
Also, these were recently utilized in the context of monitored circuits in~\cite{Beni21b}.

\subsection{Haar random code}

\begin{figure}[h!]
	\centering
	\includegraphics[width = 6cm]{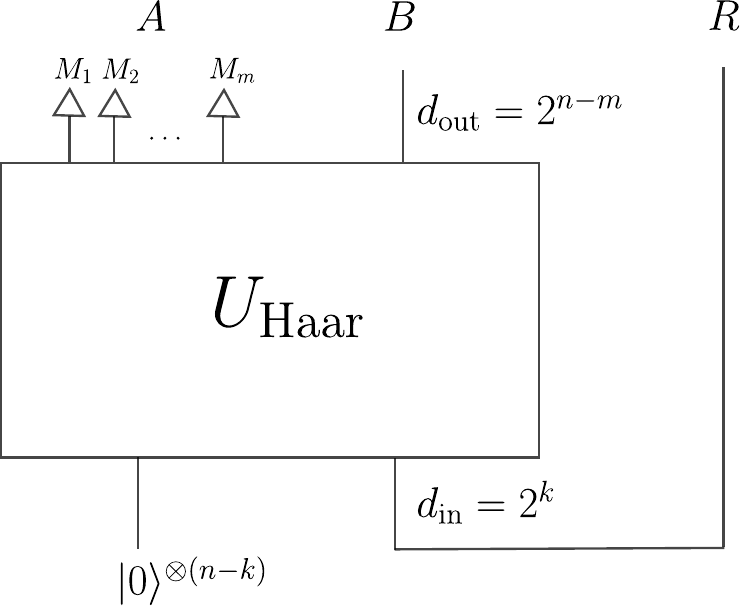}
	\caption{The Choi state of Haar random code. }
	\label{fig:Haar}
\end{figure}

Suppose that $k$ logical qubits are encoded into $n$ physical qubits through the Haar random unitary, and then $m$ physical qubits are measured in some basis as shown in Fig.~\ref{fig:Haar}. 
Note that the choice of measurement basis plays no role as the encoding is Haar random. 


The encoded information is recoverable when $B$ and $R$ in the post measurement state are nearly maximally entangled. 
Here, we are primarily interested in the average recovery fidelity. 
For this purpose, it suffices to show that the reduced density matrix $\tilde{\rho}_{R}$ is close to the maximally mixed state $\mu_{R}$.
Then, the recovery protocol from~\cite{Yoshida:2017aa} achieves a high average recovery fidelity. 

The unnormalized wavefunction after projective measurements is given by
\begin{align}
|\psi\rangle_{BR} = \langle 0|^{\otimes m}_{A}  | \Psi\rangle_{ABR}
\end{align}
where $| \Psi\rangle_{ABR}$ is the Choi state. The normalized density matrix is $\tilde{\rho}_{R} = \frac{\rho_R}{\mathrm{Tr}(\rho_R)}$.
Finally, the $2$-norm distance with $\mu_R = \frac{I_R}{d_R}$ is 
\begin{align}
\lVert \tilde{\rho}_R - \mu_R \rVert_{2}^2 = \mathrm{Tr}(\tilde{\rho}_R^2 ) - \frac{1}{d_R}.\label{eq:2norm}
\end{align}
We need to evaluate the Haar average of $\mathrm{Tr}(\tilde{\rho}_R^2 )$. We have
\begin{align}
\mathbb{E}\big(\mathrm{Tr}(\tilde{\rho}_R^2 )\big) 
= \mathbb{E}\left( \frac{\mathrm{Tr}(\rho_R^2 )}{ \mathrm{Tr}(\rho_R )^2 }  \right) 
\approx 
\frac{\mathbb{E}\big( \mathrm{Tr}(\rho_R^2 )\big) }{\mathbb{E}\big(  \mathrm{Tr}(\rho_R )^2 \big)} \label{eq:expectation}
\end{align}
where the approximation relies on the fact that variances are exponentially suppressed with respect to $n-m$ (i.e. by factors of $d_B$). 
Performing the Haar integral (see~\cite{Roberts_2017} for a review), we obtain
\begin{eqnarray}
	\mathbb{E}\big( \mathrm{Tr}(\rho_R^2 )\big)
	&=& \frac{1}{d^2_R} 
	\left( \frac{d^2_Rd_B + d_R d^2_B}{d^2_{AB}-1} - \frac{d^2_R d^2_B +d_Rd_B}{d_{AB}(d^2_{AB}-1)} \right) \cr
	\mathbb{E}\big(  \mathrm{Tr}(\rho_R )^2 \big) 
	&=& \frac{1}{d_R^2} 
	\left( \frac{d_R^2d_B^2+ d_R d_B}{d_{AB}^2 - 1} - \frac{d_Rd_B^2 + d_R^2 d_B}{d_{AB}(d_{AB}^2-1)}\right),
\end{eqnarray}
and
\begin{eqnarray}
	\frac{\mathbb{E}\big( \mathrm{Tr}(\rho_R^2 )\big)}{\mathbb{E}\big(  \mathrm{Tr}(\rho_R )^2 \big)}
	\approx
 \frac{ d_R +d_B }{d_R d_B + 1  } 
 \approx \frac{1}{d_R} \left( 1+ \frac{d_R}{d_B} \right)
	\label{Eq: renyi-2}
\end{eqnarray}
where the approximation holds for $d_{A}\gg d_{B},d_{R}$ and $d_B\gg d_R$.
This leads to 
\begin{align}
\mathbb{E}\big(\lVert \tilde{\rho}_R - \mu_R \rVert_{2}^2\big) 
\approx \frac{1}{d_B}.
\end{align}
Hence, we arrive at 
\begin{align}
\mathbb{E}\big(\lVert \tilde{\rho}_R - \mu_R \rVert_{1}^2\big) \leq d_{R}\mathbb{E}\big(\lVert \tilde{\rho}_R - \mu_R \rVert_{2}^2\big) \approx \frac{d_R}{d_B}
\end{align}
which approaches zero for $d_B\gg d_R$. 

In summary, we find that the measurement threshold of Haar random code is $p_m^{th}=1$. 
Namely, the encoded information is retained even when $n-(k+\epsilon)$ qubits are projectively measured, where the failure probability is suppressed by a factor of $2^{-\epsilon}$. 

\subsection{Haar random measurement}

\begin{figure}[h!]
\centering
\includegraphics[width = 6cm]{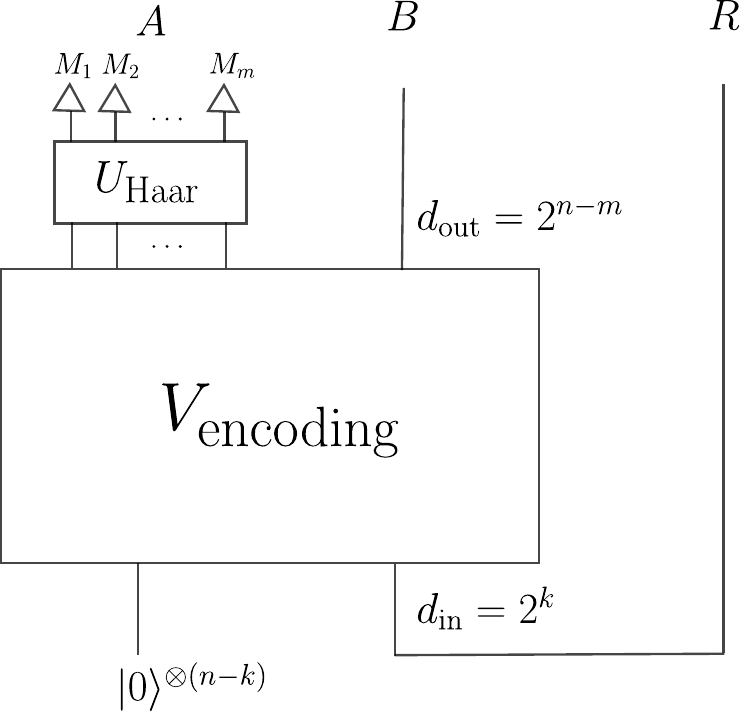}
\caption{The Choi state of arbitrary QECC code after global Haar random measurements.}
\label{fig:global haar measurement}
\end{figure}

Next, we consider an arbitrary QECC and perform a Haar random measurement on the subsystem $A$ as shown in Fig.~\ref{fig:global haar measurement}. 

Let $|\Psi\rangle_{ABR}$ be the Choi state and $\sigma_{ABR}$ be the corresponding density matrix. 
The unnormalized wavefunction after projective measurement is 
\begin{align}
|\psi\rangle_{BR} = \langle \mathrm{Haar} |_{A} |\Psi\rangle_{ABR}
\end{align}
where $|\mathrm{Haar} \rangle_{A}$ represents a Haar random state on $A$. 
This defines the unnormalized post measurement density matrix $\rho_{R}$. 
Expressing the $2$-norm as in Eq.~\eqref{eq:2norm}\eqref{eq:expectation}, our task is to evaluate $\mathbb{E}\big(\mathrm{Tr}(\tilde{\rho}_R^2 )\big)$. 
Performing the Haar integral and noting that the variances are small, we obtain 
\begin{eqnarray}
	\mathbb{E}\big( \mathrm{Tr}(\rho_R^2 )\big)
	&=&  \frac{1}{d_A(d_A+1)} \big( \mathrm{Tr}(\sigma_R^2) + \mathrm{Tr}(\sigma_B^2) \big) 
 \cr 
	\mathbb{E}\big(  \mathrm{Tr}(\rho_R )^2 \big) 
	&=& \frac{1}{d_A(d_A+1)} \big( 1 +  \mathrm{Tr}(\sigma_A^2) \big),
\end{eqnarray}
and
\begin{eqnarray}
	\mathbb{E}\big(\mathrm{Tr}(\tilde{\rho}_R^2 )\big) \approx \frac{\mathbb{E}\big( \mathrm{Tr}(\rho_R^2 )\big)}{\mathbb{E}\big(  \mathrm{Tr}(\rho_R )^2 \big)}
	\approx 
 \frac{\frac{1}{d_R} + \mathrm{Tr}(\sigma_B^2)}{1 +  \mathrm{Tr}(\sigma_A^2)} 
 \leq  \frac{1}{d_R} + \mathrm{Tr}(\sigma_B^2).
\end{eqnarray}
The encoded information will be retained in $B$ when the distance between $\tilde{\rho}_R$ and $\mu_{R}$ is small.
This occurs when $\mathbb{E}\big(\mathrm{Tr}(\tilde{\rho}_R^2 )\big)\approx \frac{1}{d_{R}}$, in particular when $\mathrm{Tr}(\sigma_B^2) \ll \frac{1}{d_R}$.
In other words, if $S^{(2)}_{B}\gg d_{R}$, the encoded information will be retained in $B$.  

It is useful to consider two particular mechanisms to achieve it. 
The first scenario is that $B$ is maximally entangled with $R$. 
In this case, the measurement on $A$ has no effects on the entanglement between $B$ and $R$, and thus, $B$ retains logical qubits after projective measurements. 
The second scenario is that $B$ is entangled with $A$, but not with $R$. 
In this case, in the original QECC, the subsystem $B$ cannot reconstruct the encoded information.
The above result suggests that, as a result of projective measurements on $A$, the encoded information becomes recoverable from $B$. 
One can interpret this as quantum teleportation of logical qubits from $A$ to $B$ by using entanglement between $A$ and $B$. 


\section{Measurement threshold in quantum gravity}\label{sec:QG}

In this section, we briefly discuss QECCs under random monitoring in the context of quantum gravity. 
The AdS/CFT correspondence can be interpreted as a QECC where the bulk degrees of freedom is holographically encoded into the boundary degrees of freedom. 
The conceptual pillar behind holographic QECCs is the entanglement wedge reconstruction~\cite{Dong_2016}.
Let us focus on the AdS$_3$/CFT$_2$ in order to illustrate the basic idea. Consider an arbitrary subsystem $A$ on the boundary and the minimal surface homologous to $A$ (i.e. the Ryu-Takayanagi (RT) surface), where  the bulk region enclosed by $A$ and the RT surface is called the entanglement wedge EA. 
The entanglement wedge reconstruction is a statement that bulk field operators in EA can be holographically reconstructed on $A$. 
Holographic tensor networks provide concrete toy model realizations of this phenomena. 

We begin by studying the fault-tolerance of the bulk quantum information $b$ at the center of the AdS space against erasure errors. 
The erasure threshold for holographic QECCs was found to be $p_e^{(th)}=\frac{1}{2}$~\cite{Pastawski_2015}.
To sketch the argument, consider a subset $A$ of qubits by choosing $pn$ qubits from the boundary with $p<\frac{1}{2}$.
We can observe that the RT surface of $A$ is not likely to enclose the bulk field $b$ at the center. 
Namely, the RT surface of the complementary subset $A^c$, which consists of $(1-p)n$ qubits with $1-p >\frac{1}{2}$, is more likely to enclose $b$. 
This suggests that the bulk quantum information $b$ is protected from erasure of qubits in $A$, and thus $p_e^{(th)}=\frac{1}{2}$. 

\begin{figure}[h!]
\includegraphics[width = 5cm]{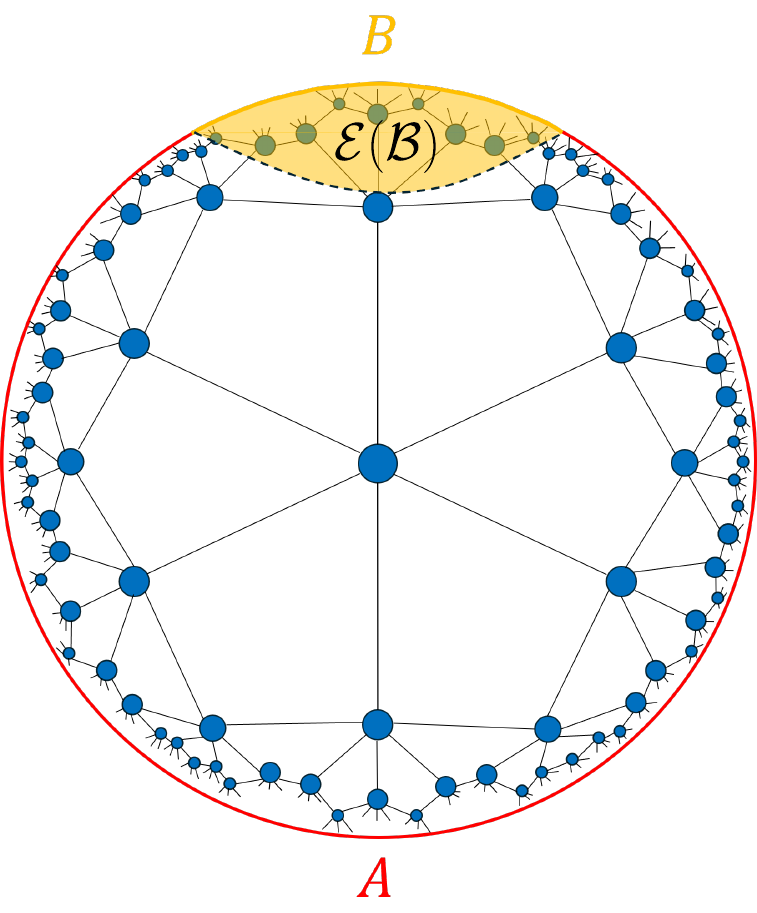}
\qquad
\includegraphics[width = 5cm]{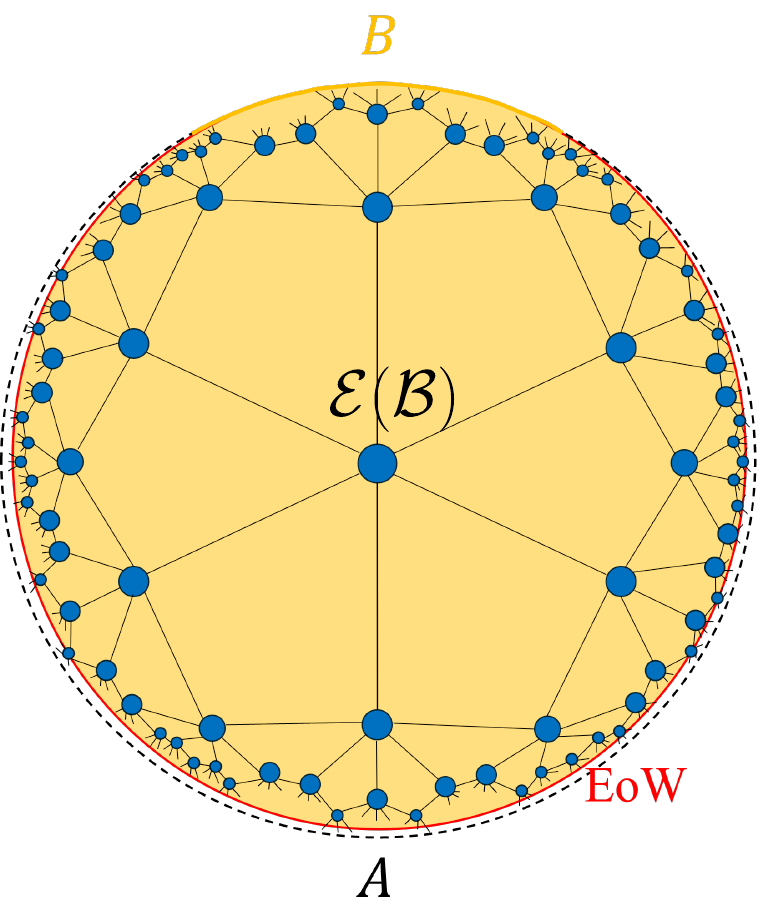}
\caption{Holographic description of measurements at the boundary subregion $A$. }
\label{fig:EoW}
\end{figure}

Next, we study the fault-tolerance against local measurements. 
The measurement threshold for holographic QECCs was argued to be $p_m^{(th)}=1$ in~\cite{Antonini_2022} by two different lines of arguments.  
First, in the AdS/CFT correspondence, it has been proposed that local measurements on the boundary are dual to insertions of End-of-World (EoW) branes in the bulk at large tension limit where branes will be situated close to the asymptotic boundary. 
Fig.~\ref{fig:EoW} shows an example of geometries with EoW branes that emerge from local projective measurements of subregion $A$.  
Suppose that we choose a subregion $A$ that consists of 99\% of the boundary qubits. 
The corresponding EoW brane will still be located near the asymptotic boundary, and the entanglement wedge of $A^c$ contains the entire bulk region, leading to $p_m^{(th)}=1$. 

One remarkable property is that we have $p_m=1$ even when the bulk field $b$ is located close to the asymptotic boundary, as long as the EoW does not encloses $b$
~\footnote{This observation ignores the $1/N$ corrections and assumes that all the bulk fields other than $b$ are negligible. Also, the maximal tension of the EoW brane is constrained by the AdS radius, see~\cite{Takayanagi_2011} for instance.}.  
This is not the case for the erasure threshold. 
Indeed, for bulk information $b$ near the boundary, erasing small subregion $A$ would suffice to lose the quantum information as $A$'s causal wedge encloses $b$. 

\section{Outlook}\label{sec:outlook}

In this paper, we have developed a theoretical framework to discuss the effect of projective measurements on QECCs. We then demonstrated that QECCs have extraordinary robustness against random local monitoring, achieving the maximal measurement threshold $p_m^{th}=1$ in many cases. 
Our work suggests that, in contrast to the conventional wisdom about projective measurements, a bit of quantum encoding helps the system to retain the memory of its initial states in the form of QECCs under projective measurements. 
Below, we provide some relevant discussions and present potential future problems. 

\textbf{Information preservation and logical measurement:}
In the 2D toric code and the 2D color code, we observed
\begin{align}
\text{Uncertainty in $(p_{\bar{X}},p_{\bar{Y}},p_{\bar{Z}})$}
\quad \Leftrightarrow \quad 
\text{$p_m^{th}=1$ for $(\alpha_X,\alpha_Y,\alpha_Z)$}.
\end{align}
We expect that this relationship may be explained from the perspective of the effective field theory descriptions induced by local random measurements. 
For the 2D toric code, a recent work pointed out that uniformly random local measurements lead to emergent Goldstone modes in the completely packed loop model with crossings~\cite{Negari23}. 

\textbf{Shadow tomography for QECC:}
The essential idea of the shadow tomography is to characterize an unknown quantum state $\rho$ by measuring it in some random basis and reconstruct $\rho$ via the inverse channel. 
When the probability distribution of the measurement basis is chosen appropriately, expectation values of various observables can be estimated by accessing only a small number of $\rho$. 
The original work~\cite{aaronson2018shadow, Huang_2020} showed that uniformly random global Pauli measurements can achieve efficient sample complexity while subsequent works~\cite{Ippoliti_2023} have made significant progress in understanding the relation between the randomness of the measurement basis and the sample efficiency. 
In the context of QECCs, we may assume that the quantum state $\rho$ of interest is unknown to us, but we hold some partial knowledge of $\rho$. 
Specifically, we already know that $\rho$ is a quantum state supported in some subspace $\mathcal{H}_{\text{code}}\subset \mathcal{H}$ that can be interpreted as the codeword subspace. 
Similar setups were considered in~\cite{Hu22, Bringewatt23}.
The shadow tomography for QECCs then requires us to characterize $\rho$ inside $\mathcal{H}_{\text{code}}$ by measuring logical operators in the random basis. 
Hence, studying the logical measurement probability $(p_{\bar{X}},p_{\bar{Y}},p_{\bar{Z}})$ allows us to estimate the sample complexity when the shadow tomography is performed by measuring physical qubits with $(p_X,p_Y,p_Z)$. 

\textbf{Clifford hierarchy:}
CSS stabilizer codes which are equipped with transversal logical gates from higher levels in the Clifford hierarchy tend to have imbalance of $X$ and $Z$ stabilizers. Namely, this leads to smaller erasure thresholds~\cite{Pastawski_2015}. 
As we have seen in the seven-qubit code and 15-qubit code, having different sets of transversal gates seems to lead to strikingly different information preservation properties. 
There may potentially be an interesting relation between transversal logical gates and the information preservation probability. 

\textbf{Decoding by measurement:}
Our work motivates a problem concerning decoding of QECCs. 
Namely, it will be interesting to ask if one can efficiently decode encoded information in QECCs by performing projective measurements in a fault-tolerant manner. 

\textbf{Topological codes:}
We have demonstrated that the 2D toric code and color code achieve $p_m^{th}=1$ under uniformly random local Pauli measurements. 
This naturally leads us to wonder if $p_m^{th}=1$ for uniformly random measurements is a generic property of 2D topological codes. 
This expectation, however, turns out to be false. 
In fact, an explicit counterexample can be constructed by considering the toric code on some lattice whose edge percolation threshold $p_{\text{perc}}$ is smaller than $\frac{1}{3}$, and thus $p_m^{th} =p_{\text{perc}} \leq  \frac{1}{3}$. 
One may formulate a weaker conjecture by observing that the 2D toric code on a square lattice has an extradordinary stability against $Y$ measurements.  
Namely, it will be interesting to ask if there exists some relative measurement frequency $(\alpha_X,\alpha_Y,\alpha_Z)$ which achieves $p_m^{th}=1$. 

\textbf{Qudit codes:}
It is also interesting to consider the information prevention condition for the qudit toric code. For a prime dimensional qudit stabilizer code, local Pauli operators are given by $X^a Z^b$ with $a,b=0,\cdots,p-1$ in terms of generalized Pauli operators. Noting that measurements in the $Q^c$ basis ($c\not=0$) are all equivalent,  independent measurement basis can be formed by $p+1$ Pauli operators, $Z, X Z^b$ with $b=0,\cdots,p-1$.
For uniformly random measurements, Pauli $X$ and $Z$ operators will be measured with probability $\frac{1}{p+1}$. 
Furthermore, for large $p$ most of measurements will occur in the $X Z^b$ ($b\not=0$) basis which are similar to $Y$ operators in the qubit toric code.
As such, we expect that the logical information will be preserved under uniformaly random measurements for a sufficiently large $p$ qudit toric code on arbitrary lattices. 

\subsection*{Acknowledgement}

We thank Alex Kubica, Amir-Reza Negari, and Shengqi Sang for useful discussions. 
Research at Perimeter Institute is supported in part by the Government of Canada through the Department of Innovation, Science and Economic Development and by the Province of Ontario through the Ministry of Colleges and Universities.

\appendix

\section{Proof of theorem~\ref{theorem:stabilizer} and~\ref{theorem:subsystem}}\label{app:proof}

We present a proof for subsystem codes since stabilizer codes are special cases of subsystem codes with $\mathcal{G}=\mathcal{S}$. We will proceed by proving i) $\Rightarrow$ ii), ii) $\Rightarrow$ iii), and iii) $\Rightarrow$ i).

\emph{Proof of }i) $\Rightarrow$ ii).
Recall that a non-trivial dressed logical operator $\ell_{\mathrm{dressed}}$ commutes with $\mathcal{S}$, and is outside $\mathcal{G}$. Hence, the statement i) suggests that elements in $\mathcal{M}$ must be either a) inside $\mathcal{G}$, or b) anti-commutes with some element in $\mathcal{S}$. 

Suppose that $\mathcal{M}\cap \mathcal{G}$ consists of $2^{m_G}$ elements with $m_G\leq m$.
Let us write $\mathcal{M}$ as follows
\begin{align}
\mathcal{M} = \langle N_1,\cdots, N_{m - m_{G}}, N_{m - m_{G}+1},\cdots, N_{m}\rangle \label{eq:generators}
\end{align}
where $N_{m - m_{G}+1},\cdots, N_{m}\in \mathcal{G}$ and $N_1,\cdots, N_{m - m_{s}}$ anti-commute with some element of $\mathcal{S}$.

Given a non-trivial bare logical operator $\ell_{\text{bare}}$, we present a generic recipe of constructing an equivalent expression $\ell'_{\text{bare}}$ that commutes with $\mathcal{M}$. 
This recipe relies on the fact that, for $N_j$ with $j=1,\cdots, m-m_G$, there exist stabilizer operators $\tilde{S}_{j}$ that anti-commute only with $N_{j}$ while commuting with all $N_{i}$ with $i\not=j$, namely
\begin{align}
N_j \tilde{S}_{i} = - (-1)^{\delta_{ij}} \tilde{S}_{i}N_j. \label{eq:step}
\end{align}
Assuming Eq.~\eqref{eq:step}, let us consider commutation relations between $\ell_{\text{bare}}$ and $N_j$. 
We have $[\ell_{\text{bare}},N_{j}]=0$ for $j=m-m_G +1,\cdots, m$ since $N_{j}$'s are stabilizer generators. 
As for $N_j$ with $j=1,\cdots, m-m_S$, if $\{\ell_{\text{bare}}, N_{j} \}=0$, we update the logical operator as $\ell'_{\text{bare}}=\ell_{\text{bare}} \tilde{S}_{j}$ so that $[\ell'_{\text{bare}}, N_{j} ]=0$.
Repeating this procedure for all $N_j$'s that anti-commute with $\ell$, one can construct $\ell'_{\text{bare}}\sim \ell_{\text{bare}}$ that commutes with $\mathcal{M}$, arriving at statement ii).

Finally, we prove the statement regarding Eq.~\eqref{eq:step}.
Let $S_{1},\cdots, S_{n-k}$ be independent stabilizer generators. 
Let $\vec{A}(N_{j})$ be a binary vector with $n-k$ entries which records the commutation relations between $N_{j}$ and $S_{i}$. 
Namely, for $i=1,\cdots, n-k$, we set 
\begin{equation}
\vec{A}(N_{j})_{i}=0,1 \qquad
N_j S_{i} = (-1)^{\vec{A}(N_{j})_{i}}S_i N_j.
\end{equation}
$\vec{A}(N_{j})$ for $j=m-m_G +1,\cdots, m$ are zero vectors because $N_{j}$'s are stabilizer generators while $\vec{A}(N_{j})$ for $j=1,\cdots, m-m_G$ must contain some nonzero entries.
One can show that $\vec{A}(N_{j})$ for $j=1,\cdots, m-m_G$  must be all independent binary vectors. 
Indeed, if one could construct a zero vector by taking linear combinations of $\vec{A}(N_{j})$'s, some product of $N_{j}$ for $j=1,\cdots, m-m_G$ would commute with all the stabilizer generators, and thus must be a dressed logical operator or a stabilizer. 
But both cases lead to contradictions, due to the construction of Eq.~\eqref{eq:generators} and the statement i), proving that $\vec{A}(N_{j})$ for $j=1,\cdots, m-m_G$ are independent from each other.
Then, by using Gaussian eliminations (see~\cite{Nielsen_Chuang} for instance), one can construct stabilizer operators $\tilde{S}_{j}$ satisfying Eq.~\eqref{eq:step}.

\emph{Proof of }ii) $\Rightarrow$ iii).
The statement ii) suggests that, for any Pauli operator $I \otimes I \otimes P_{\mathrm{bare}}$ acting on the bare reference $R_{\mathrm{bare}}$, there exists a bare logical operator $\bar{P}_{\mathrm{bare}}$ of the code that commutes with $\mathcal{M}$. 
Namely, we have $\bar{P}_{\mathrm{bare}}\otimes I_{\mathrm{gauge}} \otimes P_{\mathrm{bare}} \in \mathcal{S}_{\mathrm{Choi}}$. 
Since this operator commutes with all $M\otimes I \otimes I \in \mathcal{M}$, we have $\bar{P}_{\mathrm{bare}}\otimes I \otimes P_{\mathrm{bare}} \in \mathcal{S}'_{\mathrm{Choi}}$. 
Noting that $P_{\mathrm{bare}}$ is an arbitrary Pauli operator on $R_{\mathrm{bare}}$, one notices that $\mathcal{S}'_{R_{\mathrm{gauge}}R_{\mathrm{bare}}}$ cannot contain any non-trivial stabilizer that has non-trivial support on the reference $R_{\mathrm{bare}}$. 
Hence, due to the lemma.~\ref{lemma:subsystem}, we arrive at the statement iii). 

\emph{Proof of }iii) $\Rightarrow$ i).
We prove the contraposition. When the statement i) is not satisfied, $\mathcal{M}$ must contain a non-trivial dressed logical operator. 
Without loss of generality, assume that $\bar{P}_{\mathrm{dressed}}\in \mathcal{M}$ and $\bar{P}_{\mathrm{dressed}} \otimes P_{\mathrm{gauge}} \otimes P_{\mathrm{bare}} \in \mathcal{S}_{\mathrm{Choi}}$ where $P_{\mathrm{bare}}\not=I$. 
Since $\mathcal{M}$ is an abelian group, $\bar{P}_{\mathrm{dressed}}\in \mathcal{C}(\mathcal{M})$, and thus $\bar{P}_{\mathrm{dressed}} \otimes P_{\mathrm{gauge}} \otimes P_{\mathrm{bare}} \in \mathcal{C}(\mathcal{M})$. 
This suggests $\bar{P}_{\mathrm{dressed}} \otimes P_{\mathrm{gauge}} \otimes P_{\mathrm{bare}} \in \mathcal{S}'_{\mathrm{Choi}}$. Noting $\bar{P}_{\mathrm{dressed}} \otimes I \otimes I \in \mathcal{S}'_{\mathrm{Choi}}$, we have $I \otimes P_{\mathrm{gauge}} \otimes P_{\mathrm{bare}} \in \mathcal{S}'_{\mathrm{Choi}}$ with $P_{\mathrm{bare}}\not=I$. 
Hence, due to the lemma~\ref{lemma:subsystem}, the statement iii) is not satisfied. 

\bibliography{References.bib}

\end{document}